\documentclass[11pt,a4paper]{article}
\pdfoutput=1
\usepackage[pdftex]{graphics}
\usepackage{jheppub}
\usepackage{amsmath}
\usepackage{amssymb}

\newcommand{\be}{\begin{equation}}
\newcommand{\ee}{\end{equation}}
\newcommand{\nn}{\nonumber}
\newcommand{\bn}{\begin{enumerate}}
\newcommand{\en}{\end{enumerate}}
\newcommand{\bl}{\begin{align}}
\newcommand{\el}{\end{align}}

\def\IC{\mathbb{C}}

\def\IR{\mathbb{R}}
\def\IZ{\mathbb{Z}}

\def\CN{{\cal N}}
\def\CO{{\cal O}}

\def\CW{{\cal W}}

\def\a{\alpha}

\def\d{\delta}
\def\e{\epsilon}

\def\th{\theta}

\def\l{\lambda}
\def\m{\mu}

\def\r{\rho}

\def\s{\sigma}

\def\D{\Delta}

\def\S{\Sigma}

\def\half{\frac{1}{2}}

\def\imp{\Longrightarrow}

\def\goto{\rightarrow}

\def\tr{{\rm tr}}

\def\jmath{{j}}

\title{Klebanov-Witten flows in M-theory}

\author[a,b]{Sangmin Lee} 
\author[c]{and Sungjay Lee} 

\affiliation[a]{College of Liberal Studies, Seoul National University, Korea}
\affiliation[b]{Department of Physics and Astronomy, 
Seoul National University, Korea}
\affiliation[c]{Department of Applied Mathematics and Theoretical Physics, University of Cambridge, UK}

\abstract{
We study renormalization group flows among three dimensional 
superconformal gauge theories which closely resemble the renowned Klebanov-Witten flow in four dimensions. In the large $N$ limit, each theory appearing in the flow is holographically dual 
to M-theory on AdS$_4$ times a toric Sasaki-Einstein seven-manifold. 
The theories are obtained through the so-called flavoring method, which adds some fundamental matter fields to the dimensionally reduced Klebanov-Witten theories.
We reconfirm the matching between the gauge theories
and the dual geometries by comparing the chiral ring structure. 
As a more refined test of the flows, we compute 
the three-sphere partition function of the gauge theories. 
The square of the free energy, 
inversely proportional to the volume of the seven-manifold, 
decreases by a universal ratio 
16/27 for all flows considered in this paper.  
}

\begin{document}
\begin{flushright} {\small DAMTP-2012-50} \end{flushright}
\maketitle

\section{Introduction}

The discovery of Klebanov-Witten (KW) flow \cite{kw} 
was one of the most important landmarks in the early development of AdS/CFT correspondence. The KW flow
describes a renormalization group (RG) flow from an $\CN=2$ supersymmetric orbifold quiver gauge theory in the ultraviolet (UV) to an $\CN=1$ theory in the infrared (IR). 
The IR theory, often called the KW theory, 
was the first non-orbifold theory to be studied in the AdS/CFT context. 

The KW flow triggered rapid progress 
in several important directions. 
First, the classification of supersymmetric gravity backgrounds in terms of special holonomy manifolds was carried out in \cite{Acharya:1998db,Morrison:1998cs}. Second, 
since the KW theory was physically realized as D3 branes at the tip of 
the conifold, a toric Calabi-Yau threefold (CY$_3$) cone, 
systematic studies relating toric CY$_3$ cones to supersymmetric gauge theories was initiated in \cite{Morrison:1998cs}. 
A large number of subsequent works gradually converged to a unifying framework of the brane tiling model
\cite{Franco:2005sm,Hanany:2005ss}. 
Finally, the KW flow partly motivated the general study of RG flows in AdS/CFT. 
A notable early work includes the Pilch-Warner (PW) flow \cite{Khavaev:1998fb,Freedman:1999gp} described by a kink solution that interpolates
between the maximally supersymmetric AdS$_5\times S^5$ background 
and another AdS background for an $\CN=1$ theory 
obtained by a mass deformation. Unlike the KW flow, the PW flow 
preserves the topology of the `internal' manifold $Y$ in the 
AdS$\times Y$ background. For instance, 
for the original PW flow, the supergravity background for the IR theory is 
the same $S^5$ but with a non-standard metric and Ramond-Ramond flux. 

From early days, 
many attempts were made to carry over these important findings 
of the AdS$_5$/CFT$_4$ correspondence to the AdS$_4$/CFT$_3$ setup. 
However, progress has been severely limited due to the difficulty associated with strongly coupled infrared dynamics of M2-brane world-volume theories. 
Rather recently, the discovery of $\CN \ge 4$ supersymmetric Chern-Simons-matter theories \cite{Bagger:2006sk,Gustavsson:2007vu,Bagger:2007jr,Gaiotto:2008sd,Hosomichi:2008jd,Aharony:2008ug,Hosomichi:2008jb,Bagger:2008se,Bandres:2008ry,Schnabl:2008wj} marked a major breakthrough in the development of AdS$_4$/CFT$_3$ and revived many previous attempts. 
In particular, new methods were developed to construct a variety of $\CN=2$ theories, including 
those corresponding to M2-branes probing toric Calabi-Yau fourfold (CY$_4$) 
cones \cite{Martelli:2008si,Imamura:2008qs,hana2,Aganagic:2009zk,Benini:2009qs}. 

As for the study of RG flows, there had been gravity analysis of 
PW-like flows from early on \cite{mrg1,mrg2,mrg3,mrg4,mrg5,mrg6,mrg7}. 
More recenlty, field theory models have been given for some flows \cite{Jafferis:2011zi}, 
and the gravity analysis have also been improved \cite{Gabella:2011sg,Gabella:2012rc}. 
In contrast, the possibility of a KW-like 
flow remained an open question 
even after all the new developments. 

The main goal of this paper is to give an affirmative answer to this question 
by proposing two concrete examples of KW type flows in M-theory in the AdS$_4$/CFT$_3$ setup.  
The UV and IR CFT's appearing in the new flows of this paper are constructed by using the so-called flavoring method \cite{Aganagic:2009zk,Benini:2009qs}, which adds some fundamental matter fields to the dimensionally reduced Klebanov-Witten theories. 
Geometrically, the flavoring lifts the toric diagrams for the KW flow 
to higher dimensional ones. Once the theories are constructed, 
we can reconfirm the correspondence between the gauge theory 
and the geometry by comparing the vacuum moduli space and the chiral ring structure. 

Independently of the AdS/CFT correspondence, recently 
there has been major progress in understanding RG-flows in three dimensional 
supersymmetric gauge theories. It was shown in \cite{Kapustin:2009kz,Jafferis:2010un,Hama:2010av} that the three-sphere partition function $Z$ of $\CN=2$ supersymmetric theories can be exactly computed via localization techniques. 
The exact partition function provides a systematic and quantitative way 
to study the strongly coupled infrared dynamics of three-dimensional theories. 
In particular, Ref.~\cite{Jafferis:2010un} proposed 
that the free energy 
$F=-\text{log}|Z|$ can define a measure of the number of degrees of freedom that 
decreases monotonically along RG flows. 
See \cite{Casini:2012ei,Closset:2012vg} for possible proof of this conjectured `F-theorem'.  
When the gauge theory is AdS/CFT-dual to a toric CY$_4$, 
the free energy is related to the volume of the base $Y$ of the cone by $F \propto 1/\sqrt{{\rm Vol}(Y)}$ \cite{Herzog:2010hf,Jafferis:2011zi,Martelli:2011qj,Cheon:2011vi}. 

To test our proposal for the new RG flows at the quantum level, 
we compute the three-sphere partition function of 
the gauge theories. By extremizing the free energy with respect to 
trial $R$-charges of the matter fields, 
we find the extremal values of the $R$-charges and the value of $F$ 
that perfectly matches the expectation from the geometry. 

We also note that, for all examples of flows considered in this paper, 
the free energy decreases along the RG flow by a universal ratio, 
\be
\frac{F_{\rm IR}}{F_{\rm UV}} = \sqrt{\frac{{\rm Vol}(Y_{\rm UV})}{{\rm Vol}(Y_{\rm IR})}} = \sqrt{\frac{16}{27}} \ ,
\label{univ-ratio}
\ee
consistent with the F-theorem.
Using the localization method, we give a field theoretic proof of 
this universal ratio for a large class of theories that includes, 
but is not limited to, all theories explicitly studied in this paper. 

The universal ratio $16/27$ was first observed  in \cite{Jafferis:2011zi} 
which considered the RG flows triggered by mass deformation of three dimensional CFT's 
describing M2-branes probing CY$_3 \times \mathbb{C}$ . Here CY$_3$'s are of a special class described algebraically by $xy = z^{n_1} w^{n_2}$. 
This work provided strong hint on the existence of various PW  flows in M-theory 
whose explicit solutions were constructed recently in \cite{Gabella:2012rc}.
Furthermore, a general proof of the universal ratio was given in \cite{Gabella:2011sg} based on 
a certain scaling behavior of the gravitational free energy for the PW solutions. In the present work, we consider KW flows from CY$_3 \times \mathbb{C}$ that are qualitatively different from the PW flows. 
The two geometries dual to the UV and IR fixed points of the KW flow 
are related by a complex deformation and, in contrast to the PW flow, 
have different topology. 
Many of CY$_3$'s in our discussion are beyond the examples considered in \cite{Jafferis:2011zi}. Our field theoretical proof of the universal ratio \eqref{univ-ratio} can be regarded as AdS/CFT dual to the gravitational proof presented in \cite{Gabella:2011sg}.
See \cite{Tachikawa:2009tt} for a discussion of 
similar universal ratio in four dimensions.

This paper is organized as follows. In section 2, we review 
the basics of toric geometry. To prepare for a later comparison 
to the dual gauge theories, 
we compute explicitly the Hilbert series, which contains 
the complete information on the chiral ring in the large $N$ limit 
and also gives the volume of the Sasaki-Einstein base manifold of the toric cone.   
In section 3, after reviewing the general method of constructing 
M2 brane CFT's with fundamental matter fields via the flavoring process, 
we construct the UV and IR theories for the two basic examples of 
RG flows. To confirm the validity of the construction, 
we compute the chiral ring of the gauge theory and 
find agreement with the results from the geometry. 
In section 4, we generalize the main examples of section 3 
to infinite families by orbifolding. 
In section 5, we subject the RG flows to a more stringent test 
by computing the three-sphere partition functions. 
The partition function of the UV theory and that of IR theory 
are related in a simple manner. Using this relation, 
we give a field theoretic explanation of the universal ratio 16/27. 
Section 6 contains a brief discussion 
on the so-called brane crystal model \cite{crystal1,crystal2,crystal3}
which anticipated the KW type flows of this paper before 
the breakthrough with Chern-Simons-matter theories. 


\section{Geometry}

The $\CN=2$ superconformal gauge theories we consider in this paper 
are the world-volume theories of a stack of $N$ M2-branes near the tip of 
some toric Calabi-Yau 4-fold (CY$_4$), $X=C(Y)$ whose base (unit radius section) 
$Y$ is by definition a toric Sasakian 7-manifold.  
In what follows, we will often use the names for the base manifold $Y$, the CY$_4$ cone $X$ 
and the gauge theory dual to the geometry interchangeably.

\subsection{GLSM and toric diagram}

We follow the notation of \cite{msy1,msy2,crystal1,crystal2,crystal3} for toric geometry. 
The cone $X$ is constructed by 
the gauged linear sigma model (GLSM) \cite{Witten:1993yc}
which take  a quotient of $\IC^d$ for some $d\ge 4$.
Given some integer-valued charge matrix 
$Q_\a^I$ $(I=1, \ldots, d; \a=1, \ldots, d-4)$, the quotient 
is defined by 
\be 
X = \left\{ \{\phi_I\} \in \IC^d \left| \sum_{I=1}^d Q_\a^I |\phi_I|^2 = 0 \right.\right\} / \left(\phi_I \sim e^{Q_\a^I\th^\a} \phi_I\right).
\label{glsm}
\ee
The toric diagram is a convex polyhedron composed of a set of lattice points 
$\{v_I^i\} \in \IZ^4$ ($i=1, \cdots, 4$) satisfying 
\be 
\sum_{I=1}^d Q_\a^I v_I^i = 0.  \label{ker}
\ee
The CY condition, $\sum_I Q_\a^I=0$, enforces 
the $v_I$ to lie on the same $\IZ^3$ sublattice. 
It is customary to choose a basis to set $v_I^4=1$ for all $I$ 
and specify other three coordinates of the vertices on the $\IZ^3$ sublattice. 
Hence, the resulting toric diagram for the CY$_4$ is effectively three dimensional. 
Similarly, the toric diagram for a CY$_3$ is two dimensional. 
The toric diagram defines a solid cone 
$\D_X \equiv \{ y_i \in \IR^4 ; (v_I \cdot y) \ge 0 \mbox{ for all }I\}$ 
over which the CY$_4$ space $X$ is a $U(1)^4$ bundle.

The moduli space of K\"ahler metrics on $X$ is parameterized by the Reeb vector 
$b^i \in \IR^4$, which also defines the base of the cone by 
$Y = X \cap\{b\cdot y = 1/2\}$. 
In the basis mentioned above, the CY condition fixes $b^4=4$. 
The volume of $Y$ as an explicit function of $v_I$ and $b$ 
is known \cite{msy1, msy2}. 
To obtain the Ricci-flat metric, one minimizes the volume with respect to 
$b^{i}$ $(i=1,2,3)$ with the domain of $(b^i/4)$ being precisely the 
interior of the toric diagram. 

The CY$_4$ cone inherits a $U(1)^4$ global symmetry, $F_i$ $(i=1,2,3,4)$, 
from the GLSM construction. They all correspond to global symmetries 
of the dual gauge theory. One particular combination 
determined by the Reeb vector, $R=\half b^i F_i$, 
is dual to the superconformal $U(1)$ $R$-symmetry of the gauge theory. 

One of the most basic checks of AdS/CFT correspondence with eight supercharges 
is the comparison of chiral rings. In the gauge theory, the chiral ring 
is defined by the space of all gauge invariant monomials modded out 
by F-term conditions. In addition to the classical F-term conditions $d\CW=0$
implied by the superpotential $\CW$ of the gauge theory, the three dimensional theories considered in this paper are governed by some quantum F-term conditions. 

On the geometry side, the elements of chiral ring correspond to integer points in the cone,
$\Delta_X \cap \IZ^4 = \{ m_i \in \IR^4 ; (v_I \cdot m) \ge 0 \mbox{ for all }I\}$. 
Their $R$-charge can be computed by a simple formula $R(m) = (b\cdot m)/2$. 
In terms of the GLSM fields, the chiral ring elements are 
gauge invariant monomials of $\phi_I$. 
It is possible to assign a value of $R$-charge, $R^I$, to each GLSM field $\phi_I$. 
To determine $R^I$, one can use the correspondence between the vertices $v_I$ of the toric diagram 
and supersymmetric cycles $\S_I$ of the base manifold $Y$. Then, $R^I$ is 
proportional to the volume of $\S_I$ and satisfies $\sum_I R^I= 2$. Alternatively, 
one can compare the monomials in $\phi_I$ against the formula $R(m) = (b\cdot m)/2$ 
and deduce $R^I$. 

The generating function of the chiral ring is usually called the Hilbert series \cite{msy2,hana1}. 
\be
H_X(t) = \sum_{\{m\}}  \prod_{i=1}^4 t_i^{m_i}  
\quad \left(\{m_i\} \in \Delta_X \cap \IZ^4 \right) \,.
\ee
As explained in \cite{msy2,hana1}, 
instead of actually counting the chiral ring elements, 
one can compute the Hilbert series by a simple localization formula 
involving a triangulation of the toric diagram.  
In this paper, we will use the results of \cite{msy2,hana1} without reviewing the derivation 
of the localization formula.   
 
An important application of Hilbert series is the computation of the volume 
of the base manifold $Y$. 
The normalized volume of $Y$, defined by
\be
V_Y = \frac{{\rm Vol}(Y)}{{\rm Vol}(S^7)} = \frac{{\rm Vol}(Y)}{(\pi^4/3)} \,,
\ee
can be obtained from the Hilbert series as follows \cite{msy2,hana1}, 
\be
V_Y(b) = \lim_{\e\goto 0} \left[ \e^4 H_X(t_i=e^{-\e b_i})\right]_{b^4=4} \,.
\label{toric-volume}
\ee
By minimizing the volume with respect to $b^{i=1,2,3}$, 
we find the critical value of the Reeb vector $b_*$. 
The critical Reeb vector assigns $R$-charge $(b_*^i/2)$ to each $t_i$. 

The new KW flows in this paper is related 
to the original KW flow by dimensional reduction and flavoring process. 
Geometrically, toric diagrams for the new flows 
are related to those of the original KW flow by projection of toric diagrams. 
We will always choose the basis of the $\IZ^3$ sublattice 
such that the `vertical' projection of the toric diagrams 
for CY$_4$ reproduces those of CY$_3$ appearing in the original KW flow.

\subsection{Hilbert series and volume}

In this subsection, we compute the Hilbert series and volume 
of the four toric CY$_4$'s dual 
to the two pairs of gauge theories appearing 
in the basic examples of the RG flows.

\begin{figure}[htbp]
\begin{center}
\includegraphics[width=8.2cm]{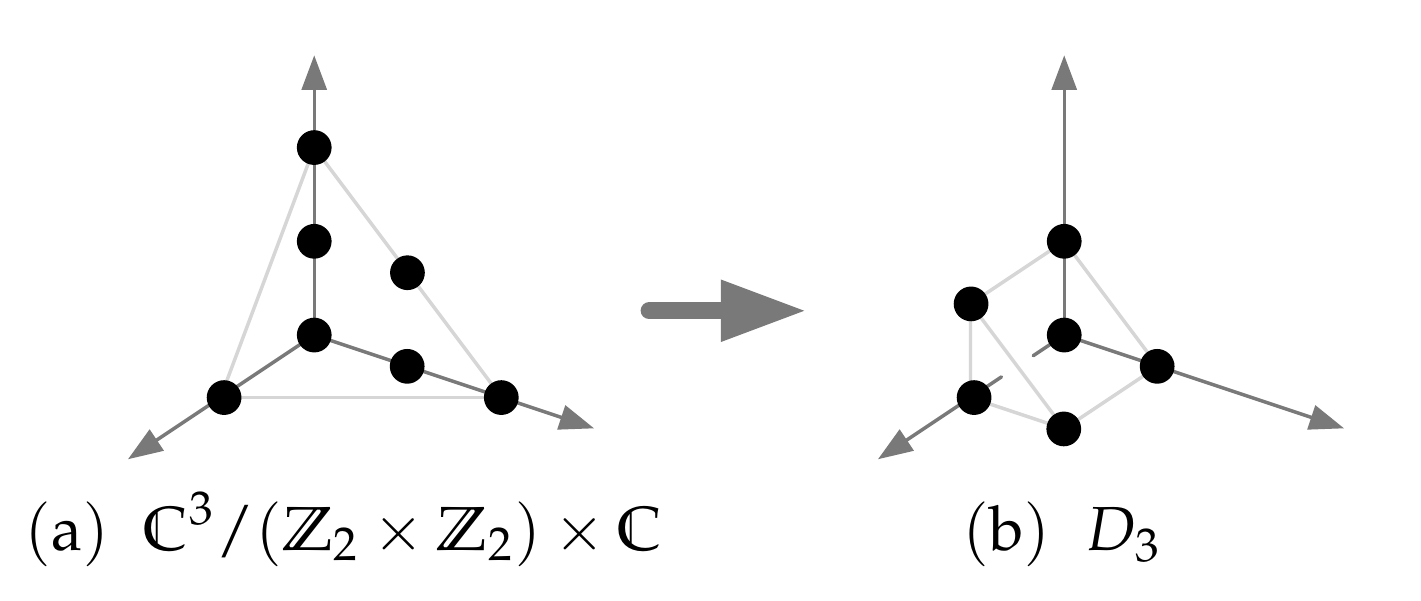}
\caption{
Toric diagrams for the flow from $\IC^3/(\IZ_2\times \IZ_2)\times \IC$ to $D_3$.
} 
\label{flow1-toric}
\end{center}
\end{figure}

\paragraph{\underline{$\IC^3/(\IZ_2\times \IZ_2)\times \IC$}} \hfill

\noindent
The toric diagram for the CY$_4$'s corresponding to the first flow is shown in Figure \ref{flow1-toric}. Note that the `vertical' projection to the 
$z=0$ plane reproduces the toric diagrams for $(\IC^2/\IZ_2)\times \IC$ 
and $C(T^{1,1})$ relevant to the original KW flow. 

Using the methods of \cite{msy2,hana1}, 
we find the Hilbert series for $\IC^3/(\IZ_2\times \IZ_2)\times \IC$, 
\be
H_{\IC^3/(\IZ_2\times \IZ_2)\times \IC}(t) = 
\frac{1+t_4/t_1}{(1-t_1)(1-t_2)(1-t_3)(1-t_4^2/t_1^2t_2t_3)} \,,
\ee
from which we can compute the normalized volume as a function of the Reeb vector,
\be
V_{\IC^3/(\IZ_2\times \IZ_2)\times \IC}(b) = \frac{2}{b_1 b_2 b_3(8-2b_1-b_2-b_3)} \,.
\label{C3-vol}
\ee
It is easy to obtain the Sasakian Reeb vector which minimizes the volume, 
\be
b_* = (1,2,2,4)\,, 
\qquad
V_{\IC^3/(\IZ_2\times \IZ_2)\times \IC}(b_*) = \frac{1}{4} \,.
\label{C3-sasa}
\ee
Note that upon making the substitution, 
\be
t_1 = s_4, \; t_2 = s_1^2, \;t_3 = s_2^2, \; 
t_4 = s_1 s_2 s_3 s_4  \\,, 
\label{C3-change}
\ee
we recover the familiar orbifold form (average over mirror images) of the Hilbert series,
\be
H_{\IC^3/(\IZ_2\times \IZ_2)\times \IC} = 
\frac{1+s_1s_2s_3}{(1-s_1^2)(1-s_2^2)(1-s_3^2)(1-s_4)}\,.
\ee
As indicated in \eqref{C3-sasa} and \eqref{C3-change}, each $s_i$ carries $R$-charge $1/2$. 

\paragraph{\underline{$D_3$}} \hfill

\noindent
The Hilbert series for $D_3$ is given by 
\be
H_{D_3}(t) = \frac{1-t_4}{(1-t_1)(1-t_4/t_1)(1-t_2)(1-t_3)(1-t_4/t_2t_3)} \,.
\ee
from which we can compute the normalized volume as a function of the Reeb vector,
\be
V_{D_3}(b)=\frac{4}{b_1(4-b_1)b_2 b_3 (4-b_2-b_3)} \,.
\label{D3-vol}
\ee
The volume-minimizing Sasakian Reeb vector respects 
the symmetry of the toric diagram,  
\be
b_* = (2,4/3,4/3,4)\,,
\qquad
V_{D_3}(b_*) = \frac{27}{64} \,.
\ee
To make the symmetries more manifest, 
we make the following change of variable, 
\be
t_1=t^3 s_1, \; t_2 = t^2 s_2, \; t_3 = t^2 s_3, \; t_4= t^6 \,,
\ee
such that \cite{hana2}
\be
H_{D_3} = \frac{1-t^6}{(1-t^3 s_1)(1-t^3/s_1)(1-t^2 s_2)(1-t^2 s_3)(1-t^2/s_2s_3)}\,.
\ee
The variable $t$ carries $R$-charge $1/3$.
The $s_i$ are fugacities for flavor symmetries. 


\begin{figure}[htbp]
\begin{center}
\includegraphics[width=7.88cm]{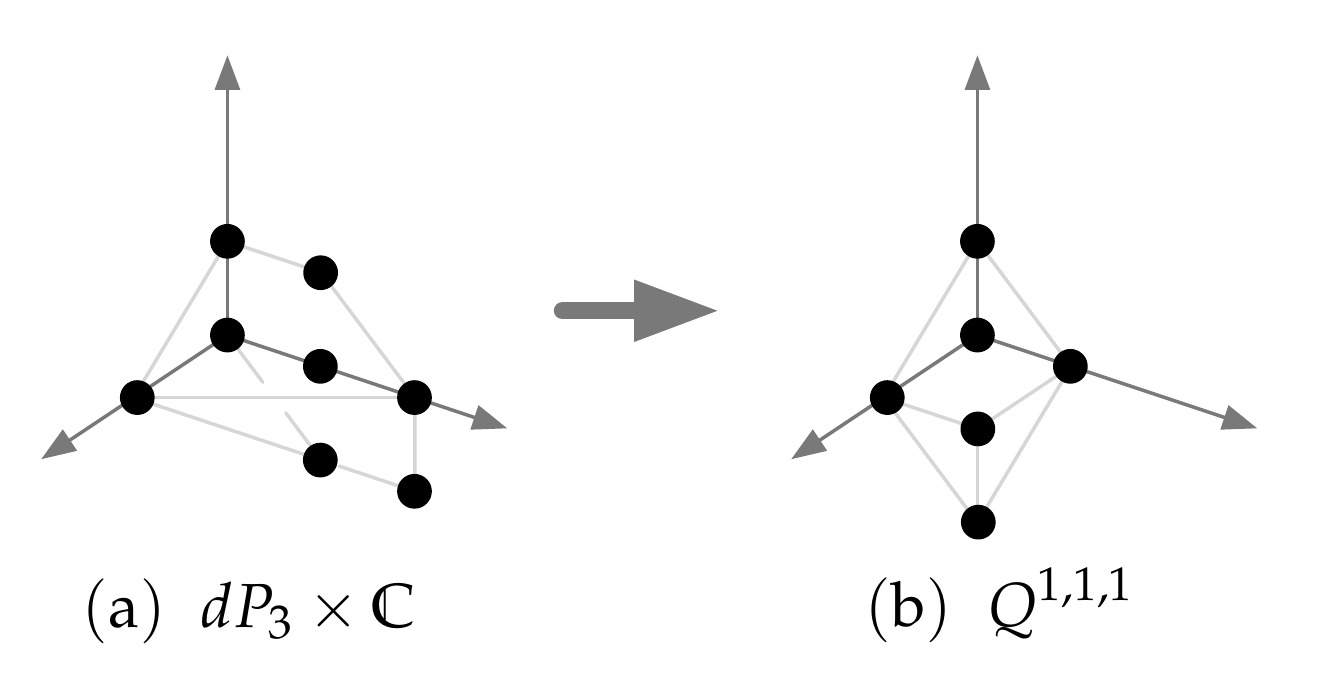}
\caption{
Toric diagrams for the flow from $dP_3\times \IC$ to $Q^{1,1,1}$.
} 
\label{flow2-toric}
\end{center}
\end{figure}

\paragraph{\underline{$dP_3\times \IC$}} \hfill

\noindent
The toric diagram for the CY$_4$'s associated to the second flow is shown in Figure \ref{flow2-toric}. 
Note again that the `vertical' projection to the 
$z=0$ plane reproduces the toric diagrams relevant to the original KW flow.

Using the methods of \cite{msy2,hana1}, 
we find the Hilbert series for $dP_3\times \IC$, 
\begin{align}
H_{dP_3\times \IC}(t) &= \frac{(1-t_4/t_1)f_{dP_3\times \IC}(t)/(t_1^4t_2t_3)}{(1-t_1) (1-t_2) (1-t_2t_3) (1-t_4/t_1t_3) (1-t_3t_4/t_1) (1-t_4^2/t_1^2t_2) (1-t_4^2/t_1^2t_2t_3)}
\,, 
\nn \\
f_{dP_3\times \IC}(t) &= t_1^4 t_2 t_3 + 2 t_1^3 t_2 t_3 t_4 + 
 2 t_1 t_2 t_3 t_4^3 + t_2 t_3 t_4^4
 \nn \\
&\quad
 - t_1^3 t_2^2 t_3 t_4 - t_1^3 t_2^2 t_3^2 t_4 - 
 t_1^2 t_2 t_4^2 - t_1^2 t_2 t_3^2 t_4^2 - t_1 t_4^3 - t_1 t_3 t_4^3 \,,
\end{align}
from which we can compute the normalized volume as a function of the Reeb vector,
\be
V_{dP_3\times \IC}(b)= 
\frac{2 (4-b_1) (32 - 16 b_1 + 2 b_1^2 + 8 b_2 - 2 b_1 b_2 - b_2^2 + 4 b_3 - 
    b_1 b_3 - b_2 b_3 - b_3^2)}{
 b_1 b_2(b_2 + b_3) (4 - b_1 + b_3) (4 - b_1 - b_3) (8 -2 b_1 -b_2)  (8 - 
    2 b_1 - b_2 - b_3)} \,.
\ee
The volume-minimizing Sasakian Reeb vector respects 
the symmetry of the toric diagram, 
\be
b_* = (1,3,0,4)\,,
\qquad
V_{dP_3\times \IC}(b_*) = \frac{2}{9} \,.
\label{dP3-sasa}
\ee
To see the symmetries of $dP_3\times \IC$ more clearly, 
we make the following change of variable, 
\be
t_1 = s_4\,, \quad 
t_2 = t s_1\,, \quad
t_3 = s_2 \,, \quad
t_4 = t s_4 \,, 
\label{dP3-change}
\ee
so that 
\begin{align}
&H_{dP_3\times \IC} = \frac{1}{(1-s_4)} H_{dP_3}(t,s_1,s_2)
\,,
\nn \\
&H_{dP_3}(t,s_1,s_2) = \left. \frac{(1-t)(1+2t-t^2\sum_{i=1}^3(s_i+1/s_i)+2t^3+t^4)}{\prod_{i=1}^3(1-ts_i)(1-t/s_i)} \right|_{s_1s_2s_3=1} .
\end{align}
As indicated in \eqref{dP3-sasa} and \eqref{dP3-change},  
the variables $(t,s_1,s_2,s_4)$ carry $R$-charges $(3/2,0,0,1/2)$. 

\paragraph{\underline{$Q^{1,1,1}$}} \hfill

\noindent
The Hilbert series of $Q^{1,1,1}$ in the basis of Figure \ref{flow2-toric}(b) 
is rather lengthy and not instructive as it 
does not show the symmetries of $Q^{1,1,1}$ manifestly. 
To recover the symmetries, we make the following change of variable, 
\be
t_1= t (s_1s_2/s_3) \,, \quad 
t_2= t (s_1/s_2 s_3) \,, \quad
t_3= s_3^2 \,, \quad
t_4 = t^2 \,.
\ee
In this new basis, the Hilbert take the simple form \cite{hana2},
\be
H_{Q^{1,1,1}} = \sum_{n=1}^\infty \chi_{n}(s_1) \chi_{n}(s_2) \chi_{n}(s_3) t^{n-1}
\,,
\ee
where 
\be
\chi_{n}(s) = \frac{s^{n}-s^{-n}}{s-s^{-1}} \,,
\ee
is the character for the $n$-dimensional representation of $SU(2)$. 
Thus we identify $s_i$ as the fugacities for the $SU(2)^3$ flavor symmetry 
of $Q^{1,1,1}$. 
The variable $t$ carries $R$-charge $1/3$.
The partial symmetries of the toric diagram in Figure \ref{flow2-toric}(b) 
uniquely determine the Sasakian Reeb vector, 
from which we find the minimal volume by using \eqref{toric-volume}, 
\be
b_* = (2,2,0,4)\,,
\qquad
V_{Q^{1,1,1}}(b_*) = \frac{3}{8} \,.
\ee

\subsection{Chiral ring \label{geo-chr}}

To prepare for comparison with gauge theories 
and later generalizations, 
we take a closer look at how 
the chiral ring can be constructed from the GLSM \eqref{glsm} 
and how its information is encoded in the Hilbert series.

\paragraph{\underline{$\IC^3/(\IZ_2\times \IZ_2)\times \IC$}} \hfill

\noindent
For the toric diagram of $\IC^3/(\IZ_2\times \IZ_2)\times \IC$ shown in Figure \ref{flow1-toric}(a), 
let $\phi_{a,b}$ denote the GLSM variable assigned 
to the vertex at $(0,a,b)$. We ignore the vertex at $(1,0,0)$ which accounts 
for the $\IC$ factor and contributes to the chiral 
ring in a trivial way. In other words, we are effectively dealing with 
a toric diagram for CY$_3$ with six GLSM fields. {}From the explicit 
value of the GLSM charges, 
\begin{equation}
\begin{array}{c|cccccc}
& \;\phi_{00}\; & \;\phi_{10}\; & \;\phi_{20}\; & \;\phi_{11}\; & \;\phi_{02}\; & \;\phi_{01}\; 
\\
\hline 
      & 0 & 1 & 2 & 1 & 0 & 0 \\
\;v\; & 0 & 0 & 0 & 1 & 2 & 1 \\
      & 1 & 1 & 1 & 1 & 1 & 1 \\
\hline
      & 1 & -2 & 1 & 0 & 0 & 0 \\
\;Q\; & 0 & 0 & 1 & -2 & 1 & 0 \\
      & 1 & 0 & 0 & 0 & 1 & -2 \\
\end{array} \,,
\end{equation}
we find that there are four elementary gauge invariant monomials 
subject to one constraint
\begin{align}
&z_1 = \phi_{00}^2 \phi_{10}\phi_{01} \,,
\quad 
z_2 = \phi_{02}^2 \phi_{01}\phi_{11} \,,
\quad 
z_3 = \phi_{20}^2 \phi_{10}\phi_{11} \,,
\nn \\
&w=\phi_{00}\phi_{10}\phi_{20}\phi_{01}\phi_{02}\phi_{11} 
\qquad \imp \qquad 
z_1 z_2 z_3 = w^2 \,.
\label{C3-glsm}
\end{align}
Thus, the chiral ring is the polynomial ring of $(z_1, z_2, z_3, w)$ modded out 
by $z_1z_2z_3=w^2$. This fact is reflected in the Hilbert series. 
For instance, the expansion, 
\begin{align}
H_{\IC^3/(\IZ_2\times \IZ_2)} &= 
\frac{1+s_1s_2s_3}{(1-s_1^2)(1-s_2^2)(1-s_3^2)} 
\nn \\
&= 1 + \underbrace{s_1^2 + s_2^2 + s_3^2}_{z_1, z_2, z_3} + \underbrace{s_1s_2s_3}_{w} + \cdots + \underbrace{s_1^2s_2^2s_3^2}_{z_1z_2z_3=w^2} + \cdots \,,
\end{align}
shows that 
there is precisely one monomial with $R$-charge 3 
that is fully invariant under the permutations of $z_1, z_2, z_3$. 

\paragraph{\underline{$D_3$}} \hfill

\noindent
For the toric diagram of $D_3$ shown in Figure \ref{flow1-toric}(b), 
let $\phi_{a,b}$ denote the GLSM variable assigned 
to the vertex at $(0,a,b)$ and similarly use $\tilde{\phi}_{a,b}$ 
for the vertex at $(1,a,b)$. There are 5 elementary gauge invariant monomials 
subject to one constraint, 
\begin{align}
&z_1 = \phi_{00}\tilde{\phi}_{00} \,, 
\quad 
z_2 = \phi_{01}\tilde{\phi}_{01} \,, 
\quad
z_3 = \phi_{10}\tilde{\phi}_{10} \,, 
\nn \\
&w = \phi_{00}  \phi_{10}  \phi_{01} \,,
\quad 
\tilde{w} = \tilde{\phi}_{00}\tilde{\phi}_{10}\tilde{\phi}_{01} 
\qquad \imp \qquad 
z_1 z_2 z_3 = w \tilde{w} \,.
\label{D3-glsm}
\end{align}
So, the chiral ring is the polynomial ring of $(z_1, z_2, z_3, w,\tilde{w})$ modded out 
by $z_1z_2z_3=w\tilde{w}$. This fact is reflected in the Hilbert series. 
For instance, the expansion, 
\begin{align}
H_{D_3} &= \frac{1-t^6}{(1-t^3 s_1)(1-t^3/s_1)(1-t^2 s_2)(1-t^2 s_3)(1-t^2/s_2s_3)}
\nn \\
&= 
1 + \underbrace{(s_2 +s_3 + 1/s_2s_3)t^2}_{z_1, z_2, z_3} 
+ \underbrace{(s_1+1/s_1)t^3}_{w,\tilde{w}} + \cdots + \underbrace{1\cdot t^6}_{z_1z_2z_3=w\tilde{w}} + \cdots\,, 
\end{align}
shows that there is precisely one monomial at $\CO(t^6)$ that 
is fully invariant under all flavor symmetries. 

We note that the two geometries $\IC^3/(\IZ_2\times\IZ_2)\times \IC$ 
and $D_3$ are related by a complex deformation;  
compare the algebraic descriptions \eqref{C3-glsm} and \eqref{D3-glsm} for the two geometries
\begin{align}
\IC^3/(\IZ_2\times\IZ_2)\times \IC\; &: \quad z_1 z_2 z_3 = w^2 \,, \quad v \;\; {\rm free}  
\nn \\
D_3 \; &: \quad z_1 z_2 z_3 = w\tilde{w} \,. 
\end{align}
We can regard the two geometries as limiting cases of a single family, 
\be
z_1 z_2 z_3 = w (w+ \e v) \,, 
\label{comp-def}
\ee
such that $\e \goto 0$ leads to $\IC^3/(\IZ_2\times\IZ_2)\times \IC$ 
and $\e=1$ with $\tilde{w} = w +v$ gives $D_3$. 

\paragraph{\underline{$dP_3\times \IC$}} \hfill

\noindent
For the toric diagram of $dP_3\times\IC$ shown in Figure \ref{flow2-toric}(a), 
let $\phi_{a,b}$ denote the GLSM variable assigned 
to the vertex at $(0,a,b)$ and ignore 
the trivial one at $(1,0,0)$. There are 7 elementary gauge invariant monomials, 
\begin{align}
&s_1 = \phi_{00}\phi_{10}\phi_{20}(\phi_{01}\phi_{11})^2 \,,
\qquad 
s_2 = \phi_{00}\phi_{10}\phi_{20}(\phi_{1,-1}\phi_{2,-1})^2 \,,
\nn \\
&t_1 = \phi_{00}\phi_{10}\phi_{20}(\phi_{01}\phi_{11})^2 \,,
\qquad \;\;
t_2 = \phi_{00}\phi_{10}\phi_{20}(\phi_{1,-1}\phi_{2,-1})^2 \,,
\nn \\
&u_1 = \phi_{10}\phi_{11}\phi_{1,-1}(\phi_{20}\phi_{2,-1})^2 \,, 
\quad 
u_2 =  \phi_{10}\phi_{11}\phi_{1,-1}(\phi_{00}\phi_{01})^2 \,,
\nn \\
& w = \phi_{00}\phi_{10}\phi_{20}\phi_{01}\phi_{11}\phi_{1,-1}\phi_{2,-1} \,,
\label{dP3-glsm}
\end{align}
subject to 9 constraints,
\begin{align}
&\qquad\quad\;\, w^2 = s_1 s_2 = t_1 t_2 = u_1 u_2, 
\nn \\
&w s_1 = t_2 u_2, \quad wt_1 = u_2 s_2, \quad wu_1 = s_2 t_2,
\nn \\
&w s_2 = t_1 u_1, \quad wt_2 = u_1 s_1, \quad wu_2 = s_1 t_1.
\label{dP3-alg1}
\end{align}
The Hilbert series correctly captures 
the structure of the chiral ring. Setting the fugacities to 1 
for simplicity, we observe that 
\begin{align}
H_{dP_3} = \frac{1+4t+t^2}{(1-t)^3} = 1+ 7 t + 19 t^2 + \cdots \,.
\end{align}
The 7 terms at $\CO(t)$ represent the elementary gauge invariant variables 
$(s_{1,2}, t_{1,2}, u_{1,2}; w)$. 
At $\CO(t^2)$, there are $\frac{7\cdot 8}{2\cdot 1}=28$ quadratic monomials in total, 
but the chiral ring relation \eqref{dP3-alg1} removes 9 of them, so we are left with 19 terms. 

\paragraph{\underline{$Q^{1,1,1}$}} \hfill

\noindent
For the toric diagram of $Q^{1,1,1}$ shown in Figure \ref{flow2-toric}(b), 
let $\phi_{a,b}$ denote the GLSM variable assigned 
to the vertex at $(0,a,b)$ and similarly use $\tilde{\phi}_{a,b}$ 
for the vertex at $(1,a,b)$. There are 8 elementary gauge invariant monomials, 
\begin{align}
&s_1 = \phi_{01}\tilde{\phi}_{00}\tilde{\phi}_{10} \,,
\quad
t_1 = \phi_{00}\tilde{\phi}_{00}\tilde{\phi}_{1,-1} \,,
\quad
u_1 =  \phi_{10}\tilde{\phi}_{10}\tilde{\phi}_{1,-1} \,,
\quad 
w_1 = \phi_{00}\phi_{10}\phi_{01} \,,
\nn \\
&s_2 = \phi_{00}\phi_{10}\tilde{\phi}_{1,-1} \,,
\quad 
t_2 = \phi_{10}\phi_{01}\tilde{\phi}_{10} \,,
\quad 
u_2 = \phi_{00}\phi_{01}\tilde{\phi}_{00} \,, 
\quad 
w_2 = \tilde{\phi}_{00} \tilde{\phi}_{10}\tilde{\phi}_{1,-1} \,, 
\label{Q111-glsm}
\end{align}
subject to 9 constraints, 
\begin{align}
&\quad\quad\quad \,w_1 w_2 = s_1 s_2 = t_1 t_2 = u_1 u_2, 
\nn \\
&w_1 s_1 = t_2 u_2, \quad w_1 t_1 = u_2 s_2, \quad w_1 u_1 = s_2 t_2,
\nn \\
&w_2 s_2 = t_1 u_1, \quad w_2 t_2 = u_1 s_1, \quad w_2 u_2 = s_1 t_1.
\label{Q111-alg1}
\end{align}
The Hilbert series correctly captures 
the structure of the chiral ring. Setting the fugacities to 1 
for simplicity, we observe that
\begin{align}
H_{Q^{1,1,1}} &= \sum_{n=1}^{\infty} n^3 t^{n-1} = 1+ 8t + 27t^2 + \cdots \,.
\end{align}
The 8 terms at $\CO(t)$ represent the elementary gauge invariant variables 
$(s_{1,2}, t_{1,2}, u_{1,2}, w_{1,2})$. 
At $\CO(t^2)$, there are $\frac{8\cdot 9}{2\cdot 1}=36$ quadratic monomials in total, 
but the chiral ring relation \eqref{Q111-alg1} removes 9 of them, so we are left with 27 terms. 

Again, the algebraic descriptions of the two geoemtries \eqref{dP3-glsm} and \eqref{Q111-glsm} are related by a complex deformation 
in a way similar to \eqref{comp-def}. 


\section{Gauge theory \label{sec-gauge}}

\subsection{M2-brane CFT with flavors}

To construct the gauge theories dual to the toric geometry 
we described in the previous section, 
we will follow the recent work \cite{Benini:2009qs} (see also \cite{Aganagic:2009zk})   which proposed a systematic method to read off the $\CN=2$ superconformal field
theories on M2-branes probing a large class of toric Calabi-Yau fourfolds CY$_4$. 
A given toric CY$_4$ can be described as a CY$_3$ fibration over a real line $\mathbb{R}=\{\sigma\}$ 
with the RR two-form field strength $F_\text{RR}$ turned on, 
where CY$_3$ can be obtained from a certain K\"ahler quotient of the CY$_4$, {\it i.e.},
CY$_3=$CY$_4//U(1)_M$. The K\"ahler moduli of CY$_3$ changes linearly as $\s$ varies. 
In terms of the 3d toric diagram for CY$_4$ with a suitable choice of $SL(3,\mathbb{Z})$ basis, such a 
K\"ahler quotient can be understood as a vertical projection down to the 2d toric diagram for CY$_3$ on the $z=0$ plane.  
Performing the Kaluza-Klein (KK) reduction along the circle $U(1)_M$, 
the M2-branes at the tip of the CY$_4$ singularity can be reduced to D2-branes probing the CY$_3$
with the two-form flux turned on. One can construct a low-energy quiver gauge theory living on the  
D2-branes using standard methods such as the brane-tiling model \cite{Franco:2005sm,Hanany:2005ss}. 
One key feature in the type IIA background is that the two-form flux $F_\text{RR}$ can induce 
the Chern-Simons coupling to the low-energy quiver gauge theory. 

We will consider in this paper some examples where the M-theory circle  
become degenerate in the KK reduction. 
In particular, we are interested in the case where $U(1)_M$ action has fixed loci 
of codimension two which are non-compact. In terms of the 3d toric diagram, the degeneration
happens when two adjacent external toric vertices are projected down to the same external point in 
the 2d toric diagram. In the type IIA background, 
it leads to D6-branes wrapping a toric divisor, a non-compact four-cycle, corresponding to the
external point in the 2d toric diagram. Let $X_\alpha$ denote a bi-fundamental matter field 
in the quiver gauge theory associated to the toric divisor. Then, one can 
argue that D6-branes introduce to the quiver gauge theory 
$n_I$ pairs of flavors, coupled to X$_I$ via the superpotential 
\begin{align}
  \CW_\text{flavor} = \sum_{a=1}^{n_\a} p^a_\a X_\a q^a_\a \ .
  \label{flavorsuperpotential}
\end{align}
We refer to the above process as flavoring the gauge theory. 

It is possible to reverse the flavoring process and construct the CY$_4$ background in M-theory from a flavored gauge theory 
by analyzing the vacuum moduli space. It often turns out that (diagonal) 
monopole operators $T^{(n)}$ play a key role in identifying the quantum moduli space of the quiver gauge theory. 
The monopole operator $T^{(n)}$ carries the same flux $n$ for 
all diagonal $U(1)$ gauge groups in the quiver together with electric charges $(nk_1,\ldots,
nk_G)$ where $k_I$ denotes the Chern-Simons level for each $U(1)$ gauge group.

As shown in \cite{Benini:2009qs}, for the flavored quiver gauge theories, 
the monopole operators becomes charged under both gauged and global $U(1)$ symmetry groups
via quantum corrections
\begin{align}
  \d Q[ T^{(n)} ]= - \frac{|n|}{2} \sum_{f} Q_f\ ,   
  \label{quantumcharge}
\end{align}
where $Q_f$ denote the $U(1)$ charge for matter fermions. 
Based on (\ref{flavorsuperpotential}) and (\ref{quantumcharge}), 
one can show that the monopole operator $T^{(n)}$ carries flavor $U(1)$ charges
\begin{align}
  Q[T^{(n)}] = |n| \sum_{\a} n_\a Q[X_\a] \ ,
\end{align}
a $U(1)_R$ charge
\begin{align}
  R[T^{(n)}] = |n| \sum_{\a} n_\a R[X_\a] \ ,
\end{align}
and gauge charges $g_a$ ($a=1,2,\ldots,G$)
\begin{align}
  g_a [T^{(n)}] = nk_a + |n| \sum_{\a} n_\a g_a[X_\a] \ .
\end{align}
Those quantum-mechanically generate charges strongly implies that 
the following holomorphic quantum relation, known as `quantum F-relation', 
should hold
\begin{align}
  T^{(n)} T^{(-n)} = \left( \prod_\a X_\a^{h_\a} \right)^{|n|}\ .
\end{align}
As a consequence, the simplest monopole operators $T\equiv T^{(1)}$ 
and $\tilde T \equiv T^{(-1)}$ parametrizes one extra dimension beyond 
those for CY$_3$, needed for the CY$_4$ background in M-theory.

\subsection{Klebanov-Witten flow \label{kw-ex}}

The UV theory of the original KW flow has the superpotential,
\be
\CW_{\rm UV} = {\rm tr}\left[ \Phi(A_1B_2-A_2B_1) - \widetilde{\Phi} (B_2A_1-B_1A_2)\right] \,.
\ee
The flow is triggered by the addition of the relevant operator,
\be
\CW_{\rm def} = \frac{m}{2} {\rm tr}(\widetilde{\Phi}^2-\Phi^2) \,, 
\label{W-def-kw}
\ee
which is the field theoretic counterpart 
of the complex deformation of geometry discussed in section \ref{geo-chr}. 
Integrating out $\Phi$, $\widetilde{\Phi}$, 
we obtain the superpotential of the IR theory, 
\be
\CW_{\rm IR} = \frac{1}{m} {\rm tr}(A_1B_1A_2B_2-A_1B_2A_2B_1) \,.
\ee

The key idea in the construction of the gauge theories in this paper is that, 
for both the UV theory and the IR theory of the KW flow, 
we can apply dimensional reduction to three dimensions 
and addition of fundamental chiral multiplets according to the 
rules of `flavoring' explained earlier. 
Our main claim is that there exists an RG flow between 
the resulting UV and IR theories in three dimensions.

\begin{figure}[htbp]
\begin{center}
\includegraphics[width=10cm]{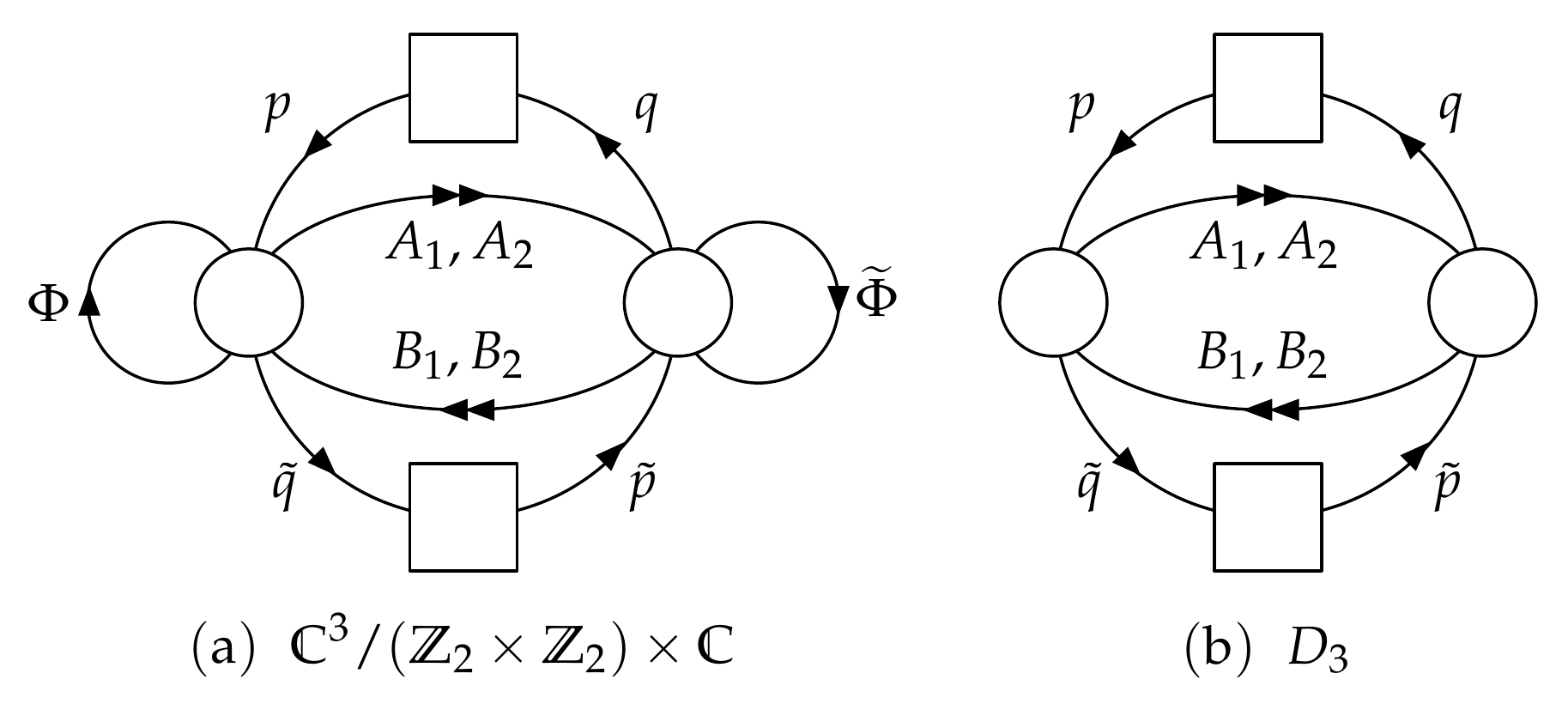}
\caption{
Quiver diagrams for the flow from $\IC^3/(\IZ_2\times \IZ_2)\times \IC$ to $D_3$.
} 
\label{flow1-quiver}
\end{center}
\end{figure}

For the flow from $\IC^3/(\IZ_2\times \IZ_2)\times \IC$ to 
$D_3$, we put `flavors' on $A_1$ and $B_1$ 
by adding the superpotential term,  
$\CW_{\rm flavor} = pA_1 q +\tilde{p} B_1 \tilde{q}$, as depicted in Figure \ref{flow1-quiver}. 
Then, we begin with the UV theory with 
\be
\CW_{\rm UV} = {\rm tr}\left[ \Phi(A_1B_2-A_2B_1) - \widetilde{\Phi} (B_2A_1-B_1A_2)\right] + p A_1 q + \tilde{p} B_1 \tilde{q}  \,,
\ee
and trigger the KW flow to end up with the IR theory with 
\be
\CW_{\rm IR} = \frac{1}{m} {\rm tr}(A_1B_1A_2B_2-A_1B_2A_2B_1) + p A_1 q + \tilde{p} B_1 \tilde{q}  \,,
\ee
Both for the UV and IR theories, 
the quantum F-term relation $T\widetilde{T} = A_1 B_1$ plays a crucial role.

\begin{figure}[htbp]
\begin{center}
\includegraphics[width=10.1cm]{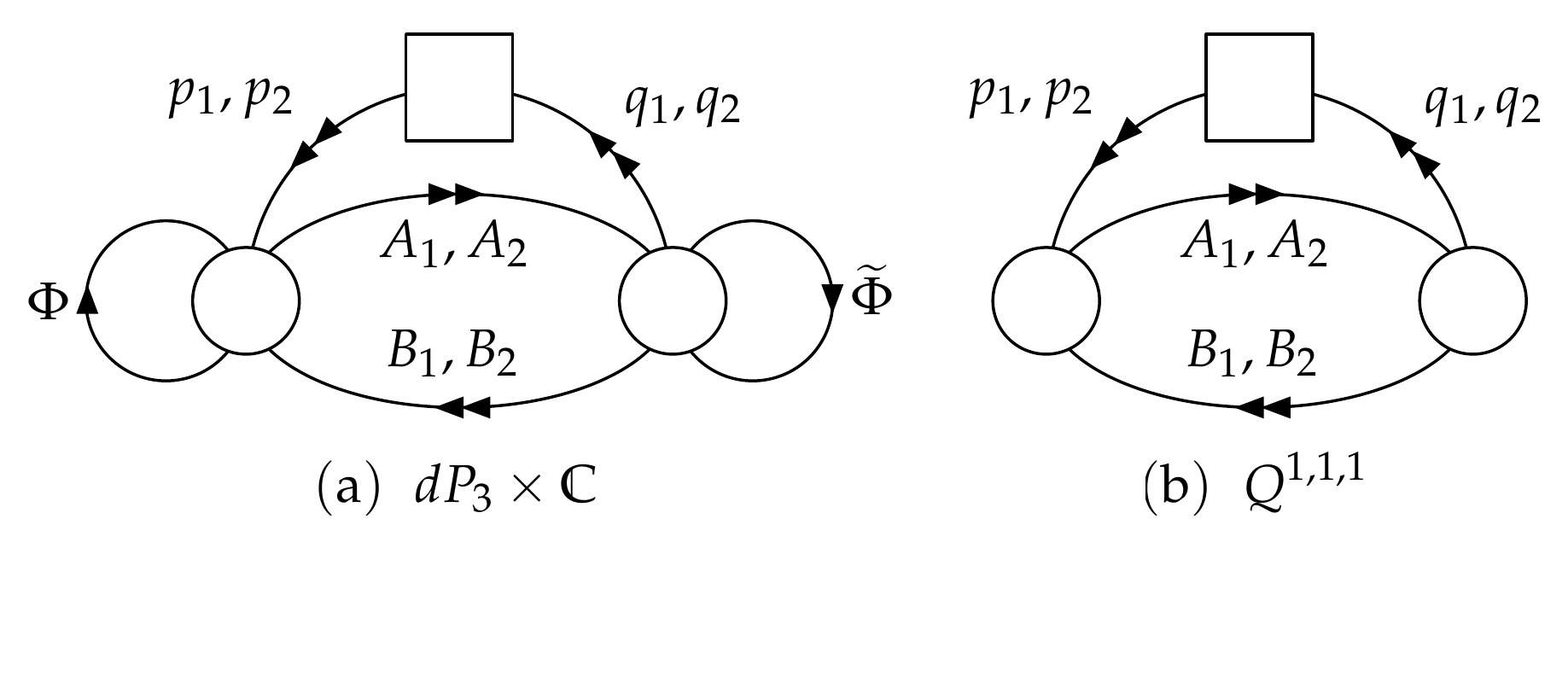}
\vskip -1.2cm
\caption{
Quiver diagrams for the flow from $dP_3\times \IC$ to $Q^{1,1,1}$.
} 
\label{flow2-quiver}
\end{center}
\end{figure}

Similarly, for the flow from $dP_3\times \IC$ to 
$Q^{1,1,1}$, 
we put `flavors' on $A_1$ and $A_2$ by 
adding the term, 
$\CW_{\rm flavor} = p_1 A_1 q_1 + p_2 A_2 q_2$ 
as depicted in Figure \ref{flow2-quiver}. 
The resulting UV theory has the superpotential, 
\be
\CW_{\rm UV} = {\rm tr}\left[ \Phi(A_1B_2-A_2B_1) - \widetilde{\Phi} (B_2A_1-B_1A_2)\right] + p_1 A_1 q_1 + p_2 A_2 q_2 \,,
\ee
and the IR theory has
\be
\CW_{\rm IR} = \frac{1}{m} {\rm tr}(A_1B_1A_2B_2-A_1B_2A_2B_1) + 
p_1 A_1 q_1 + p_2 A_2 q_2 \,.
\ee
The quantum F-term relation reads $T\widetilde{T} = A_1 A_2$.

\subsection{Chiral ring and R-charge}

It is straightforward to show that the chiral 
ring spectrum of the four theories above 
matches those from the corresponding geometries.

\paragraph{
\underline{$\IC^3/(\IZ_2\times \IZ_2)\times \IC$}} \hfill

\noindent
The gauge invariant generators of the chiral ring are
\be
z_1 = T, \quad z_2 = \widetilde{T}, \quad z_3 = A_2B_2 , \quad 
w = A_1 B_2 = A_2 B_1, \quad 
v = \Phi = \widetilde{\Phi} .
\label{C3-gauge}
\ee
The quantum F-term relation $T\widetilde{T} = A_1B_1$ implies 
that $z_1 z_2 z_3 = w^2$, which captures the $\IC^3/(\IZ_2\times \IZ_2)$ orbifold geometry. The remaining variable $v$ parametrizes 
the $\IC$ factor. 

It is instructive to compare the chiral ring between the gauge theory and the geometry. Recall from section \ref{geo-chr} 
the GLSM description of the chiral ring, 
\begin{align}
&z_1 = \phi_{00}^2 \phi_{10}\phi_{01} \,,
\quad 
z_2 = \phi_{02}^2 \phi_{01}\phi_{11} \,,
\quad 
z_3 = \phi_{20}^2 \phi_{10}\phi_{11} \,,
\quad
w=\phi_{00}\phi_{10}\phi_{20}\phi_{01}\phi_{02}\phi_{11} \,.
\label{C3-glsm-copy}
\end{align}
Ignoring the trivial $\IC$ factor and 
comparing \eqref{C3-gauge} and \eqref{C3-glsm-copy}, 
we find the correspondence between the matter fields in gauge theory and the GLSM variables,
\be
A_1 = \phi_{00}\phi_{01}\phi_{02}\,,
\quad
A_2 = \phi_{20}\,,
\quad
B_1 = \phi_{00}\phi_{01}\phi_{02}\phi_{10}\phi_{11}\,,
\quad
B_2 = \phi_{10}\phi_{11}\phi_{20}\,.
\label{C3-cor}
\ee
Using the $R$-charge of GLSM variables computed earlier and 
the correspondence \eqref{C3-cor}, we find the $R$-charge of the matter fields, 
\begin{equation}
\begin{array}{c|ccccccc}
& \Phi & \widetilde{\Phi} & A_1 & A_2 & B_1 & B_2 & \;\; p, q, \tilde{p}, \tilde{q}\;\; \\
\hline 
\; R \; & \;1/2\; & \;1/2\; & \;1\; & \;1/2\; & \;1\; & \;1/2\; & \;1/2\; 
\end{array} \,.
\end{equation}
We will show in section \ref{ZF} that the quantum computation 
of the three-sphere partition function 
reproduces exactly the same $R$-charge spectrum. 

\paragraph{\underline{$D_3$}} \hfill

\noindent
The gauge invariant generators of the chiral ring are
\be
z_1 = T, \quad z_2 = \widetilde{T}, \quad z_3 = A_2B_2 , \quad 
w = A_1 B_2, \quad \tilde{w} = A_2 B_1. 
\label{D3-gauge}
\ee
They satisfy the F-term relation $z_1 z_2 z_3 = w \tilde{w}$, 
which describes the $D_3$ geometry. 
Recall the GLSM description of the chiral ring, 
\begin{align}
&z_1 = \phi_{00}\tilde{\phi}_{00} \,, 
\quad 
z_2 = \phi_{01}\tilde{\phi}_{01} \,, 
\quad
z_3 = \phi_{10}\tilde{\phi}_{10} \,, 
\quad
w = \phi_{00}  \phi_{10}  \phi_{01} \,,
\quad 
\tilde{w} = \tilde{\phi}_{00}\tilde{\phi}_{10}\tilde{\phi}_{01} \,.
\label{D3-glsm-copy}
\end{align}
Comparing \eqref{D3-gauge} and \eqref{D3-glsm-copy}, we find
\be
A_1 = \phi_{00}\phi_{01} \,,
\quad
A_2 = \tilde{\phi}_{10}\,,
\quad
B_1 = \tilde{\phi}_{00}\tilde{\phi}_{01}\,,
\quad
B_2 = \phi_{10} \,,
\ee
which implies the following $R$-charge assignments for the matter fields,
\begin{equation}
\begin{array}{c|ccccccc}
& \Phi & \widetilde{\Phi} & A_1 & A_2 & B_1 & B_2 & \;p, q, \tilde{p}, \tilde{q} \;
\\
\hline 
\;R\; & \;1\; & \;1\; & \;2/3\; & \;1/3\; & \;2/3\; & \;1/3\; & \;2/3\;    
\end{array} \,.
\end{equation}

\paragraph{\underline{$dP_3\times \IC$}} \hfill

\noindent
The gauge invariant generators of the chiral ring are
\begin{align}
&s_1 = TB_1, \quad s_2 = \widetilde{T}B_2, \quad 
t_1 = \widetilde{T}B_1, \quad t_2 = T B_2, \quad  
\nn \\
&u_1 = A_2 B_2, \quad u_2 = A_1 B_1,  \quad
w = A_1 B_2 = A_2 B_1, \quad 
v = \Phi = \widetilde{\Phi}.
\label{dP3-gauge}
\end{align}
The quantum F-term relation, $T\widetilde{T}=A_1A_2$, 
implies the following chiral ring relations, 
\begin{align}
&\qquad\quad\;\, w^2 = s_1 s_2 = t_1 t_2 = u_1 u_2, 
\nn \\
&w s_1 = t_2 u_2, \quad wt_1 = u_2 s_2, \quad wu_1 = s_2 t_2,
\nn \\
&w s_2 = t_1 u_1, \quad wt_2 = u_1 s_1, \quad wu_2 = s_1 t_1.
\label{dP3-alg2}
\end{align}
Recall the GLSM description of the chiral ring, 
\begin{align}
&s_1 = \phi_{00}\phi_{10}\phi_{20}(\phi_{01}\phi_{11})^2 \,,
\qquad 
s_2 = \phi_{00}\phi_{10}\phi_{20}(\phi_{1,-1}\phi_{2,-1})^2 \,,
\nn \\
&t_1 = \phi_{00}\phi_{10}\phi_{20}(\phi_{01}\phi_{11})^2 \,,
\qquad \;\;
t_2 = \phi_{00}\phi_{10}\phi_{20}(\phi_{1,-1}\phi_{2,-1})^2 \,,
\nn \\
&u_1 = \phi_{10}\phi_{11}\phi_{1,-1}(\phi_{20}\phi_{2,-1})^2 \,, 
\quad 
u_2 =  \phi_{10}\phi_{11}\phi_{1,-1}(\phi_{00}\phi_{01})^2 \,,
\nn \\
& w = \phi_{00}\phi_{10}\phi_{20}\phi_{01}\phi_{11}\phi_{1,-1}\phi_{2,-1} \,.
\label{dP3-glsm-copy}
\end{align}
Comparing \eqref{dP3-gauge} and \eqref{dP3-glsm-copy}, we find
\be
A_1 = \phi_{00}\phi_{01} ,
\;
A_2 = \phi_{20}\phi_{2,-1} ,
\;
B_1 = \phi_{00}\phi_{01}\phi_{1,-1}\phi_{10}\phi_{11},
\;
B_2 = \phi_{1,-1}\phi_{10}\phi_{11}\phi_{20}\phi_{2,-1},
\ee
which implies the following $R$-charge assignments of the matter fields, 
\begin{equation}
\begin{array}{c|ccccccc}
& \Phi & \widetilde{\Phi} & A_1 & A_2 & B_1 & B_2 & p, q, \tilde{p}, \tilde{q} \\
\hline 
\;R\; & \;1/2\; & \;1/2\; & \;1\; & \;1\; & \;1/2\; & \;1/2\; & \;1/2\; 
\end{array} \,.
\end{equation}

\paragraph{\underline{$Q^{1,1,1}$}} \hfill

\noindent
The gauge invariant generators of the chiral ring are
\begin{align}
&s_1 = TB_1, \quad s_2 = \widetilde{T}B_2, \quad 
t_1 = \widetilde{T}B_1, \quad t_2 = T B_2, \quad  
\nn \\
&u_1 = A_2 B_2, \quad u_2 = A_1 B_1,  \quad
w_1 = A_1 B_2, \quad w_2 = A_2 B_1.
\label{Q111-gauge}
\end{align}
The chiral ring relations read 
\begin{align}
&\quad\quad\quad \,w_1 w_2 = s_1 s_2 = t_1 t_2 = u_1 u_2, 
\nn \\
&w_1 s_1 = t_2 u_2, \quad w_1 t_1 = u_2 s_2, \quad w_1 u_1 = s_2 t_2,
\nn \\
&w_2 s_2 = t_1 u_1, \quad w_2 t_2 = u_1 s_1, \quad w_2 u_2 = s_1 t_1.
\label{Q111-alg2}
\end{align}
Recall the GLSM description of the chiral ring, 
\begin{align}
&s_1 = \phi_{01}\tilde{\phi}_{00}\tilde{\phi}_{10} \,,
\quad
t_1 = \phi_{00}\tilde{\phi}_{00}\tilde{\phi}_{1,-1} \,,
\quad
u_1 =  \phi_{10}\tilde{\phi}_{10}\tilde{\phi}_{1,-1} \,,
\quad 
w_1 = \phi_{00}\phi_{10}\phi_{01} \,,
\nn \\
&s_2 = \phi_{00}\phi_{10}\tilde{\phi}_{1,-1} \,,
\quad 
t_2 = \phi_{10}\phi_{01}\tilde{\phi}_{10} \,,
\quad 
u_2 = \phi_{00}\phi_{01}\tilde{\phi}_{00} \,, 
\quad 
w_2 = \tilde{\phi}_{00} \tilde{\phi}_{10}\tilde{\phi}_{1,-1} \,, 
\label{Q111-glsm-copy}
\end{align}
Comparing \eqref{Q111-gauge} and \eqref{Q111-glsm-copy}, we find
\be
A_1 = \phi_{00}\phi_{01} \,,
\quad
A_2 = \tilde{\phi}_{10}\tilde{\phi}_{1,-1}\,,
\quad
B_1 = \tilde{\phi}_{00}\,,
\quad
B_2 = \phi_{10} \,,
\ee
which implies the following $R$-charge assignments of the matter fields, 
\begin{equation}
\begin{array}{c|ccccccc}
& \Phi & \widetilde{\Phi} & A_1 & A_2 & B_1 & B_2 & p, q, \tilde{p}, \tilde{q} \\
\hline 
\;R\; & \;1\; & \;1\; & \;2/3\; & \;2/3\; & \;1/3\; & \;1/3\; & \;2/3\; 
\end{array} \,.
\end{equation}

\section{Generalization to Orbifolds \label{sec-orbi}}

\subsection{Orbifolding the 4d flows }

\paragraph{\underline{$\IC^2/\IZ_{2n}\times \IC \goto T^{1,1}/\IZ_n$}} \hfill

\begin{figure}[htbp]
\begin{center}
\includegraphics[width=8cm]{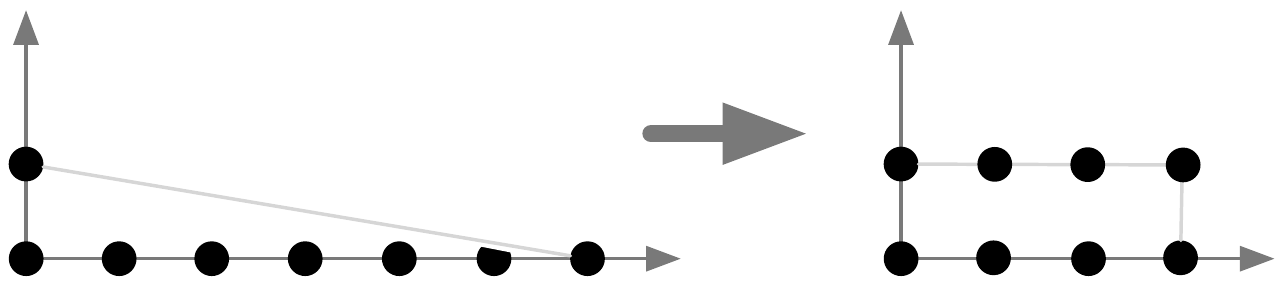}
\caption{Toric diagrams for the flow from $\IC^2/\IZ_{2n}\times \IC$ to $T^{11}/\IZ_n$
} 
\label{4dflow1-orbi-toric}
\end{center}
\end{figure}

\noindent
The original KW flow admits a simple generalization to an infinite family of  orbifolds. 
On the geometry side, the orbifolded flows relate $\IC^2/\IZ_{2n}\times \IC$ to 
$T^{1,1}/\IZ_n$. The toric diagrams for the two orbifolds are depicted in Figure \ref{4dflow1-orbi-toric}. 

\begin{figure}[htbp]
\begin{center}
\includegraphics[width=12cm]{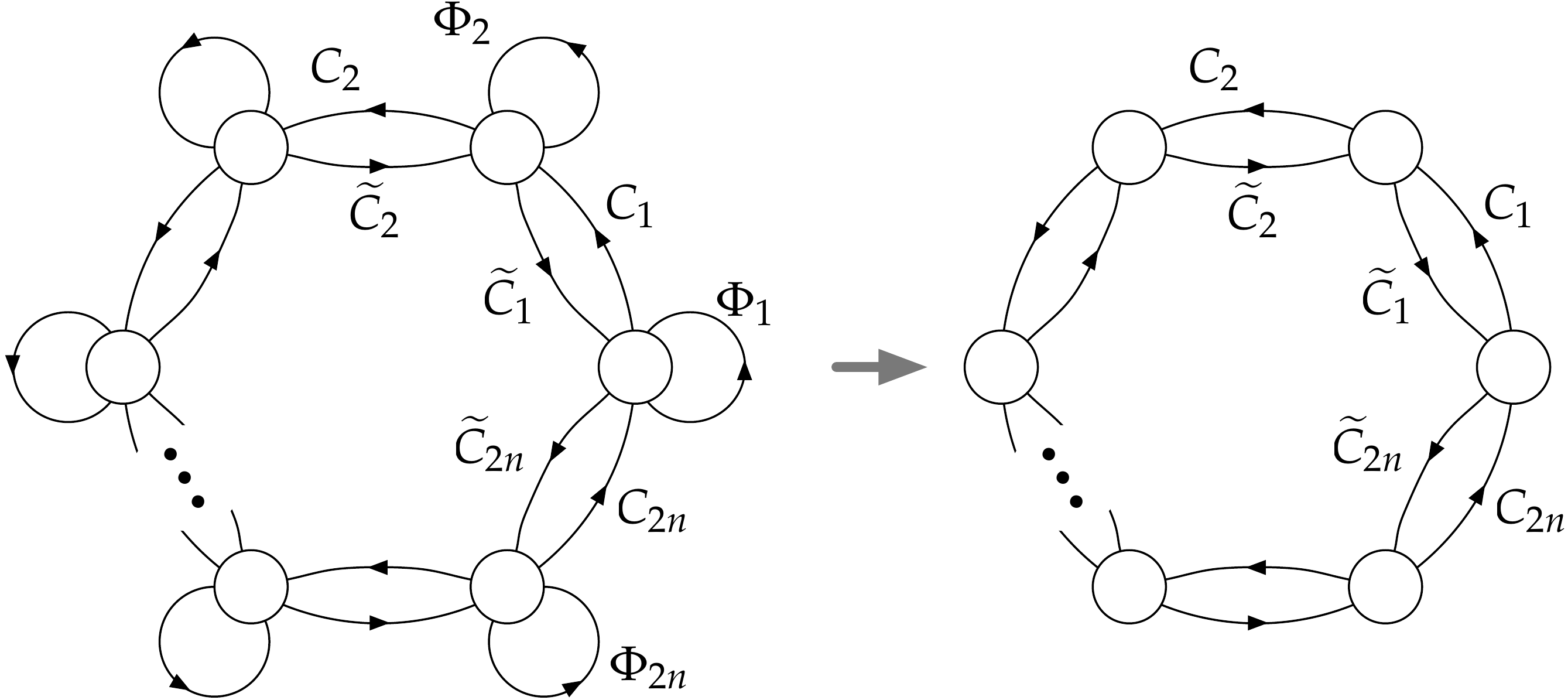}
\caption{Toric diagrams for the flow from $\IC^2/\IZ_{2n}\times \IC$ to $T^{11}/\IZ_n$
} 
\label{4dflow1-orbi-quiver}
\end{center}
\end{figure}

\noindent
On the gauge theory side, the orbifold action 
results in `necklace' quiver theories whose quiver diagrams are shown in Figure \ref{4dflow1-orbi-quiver}. 

\noindent
The UV theory of the orbifolded KW flow has the superpotential,
\be
\CW_{\rm UV} = \sum_{I=1}^{2n} \tr\left[ \Phi_I (C_I \widetilde{C}_I - \widetilde{C}_{I-1} C_{I-1}) \right] \,.
\ee
The flow is triggered by the addition of the relevant operator,
\be
\CW_{\rm def} = \frac{m}{2} \sum_{I=1}^{2n} (-1)^{I}\tr( \Phi_I^2 ) \,.
\label{W-def-quiver}
\ee
Integrating out the adjoint fields, we obtain the superpotential of the IR theory,
\be
\CW_{\rm IR} = \frac{1}{m} \sum_{I=1}^{2n} (-1)^{I} \tr\left( C_I \widetilde{C}_I \widetilde{C}_{I-1} C_{I-1} \right) \,.
\ee
To compare with the notations of the original ($n=1$) KW flow discussed in section \ref{kw-ex}, 
we identify the fields as follows
\be
(\Phi, \widetilde{\Phi}, A_1, A_2, B_1, B_2)_{\rm there} \goto 
(\Phi_1, \Phi_2, C_1, \widetilde{C}_2, C_2, \widetilde{C}_1 )_{\rm here} \,.
\ee

To see the matching between the quiver theory and the orbifold geometry, 
we examine the chiral ring. In the UV theory, enumerating all elementary gauge invariant 
operators from the (abelian) field theory and modding out by the F-term conditions, 
we find the generators of the chiral ring, 
\begin{align}
x = \prod_{I=1}^{2n} C_I \,, 
\quad
y = \prod_{I=1}^{2n} \widetilde{C}_I \,, 
\quad
z = C_1 \widetilde{C}_1 = \ldots  =  C_{2n} \widetilde{C}_{2n} \,, 
\quad 
v = \Phi_1 = \ldots = \Phi_{2n} \,.
\label{C2n-gauge}
\end{align}
They satisfy one constraint equation, 
\be
xy=z^{2n}\,, \quad v \;\; {\rm free} \,,
\label{C2n}
\ee
which describes $\IC^2/\IZ_{2n}\times \IC$ algebraically. 
We can see the same chiral ring structure from 
the toric diagram. We ignore the trivial $\IC$ factor 
and let $\phi_k$ denote the GLSM fields associated to vertices 
on the `$x$-axis' in the diagram on the left in Figure \ref{4dflow1-orbi-toric}. 
Then, the gauge invariant monomials,  
\begin{align}
x = \prod_{k=0}^{2n} (\phi_k)^{2n-k} \,,
\quad
y = \prod_{k=0}^{2n} (\phi_k)^k\,, 
\quad 
z  = \prod_{k=0}^{2n} \phi_k \,,
\label{C2n-glsm} 
\end{align}
satisfy the same algebraic relation \eqref{C2n}. 
Comparing \eqref{C2n-gauge} and \eqref{C2n-glsm}, 
we find the correspondence between gauge theory variables and GLSM variables, 
\begin{align}
C_j = \prod_{k=0}^{j-1} \phi_k \,,
\quad 
\widetilde{C}_j = \prod_{k=j}^{2n} \phi_k \,.
\label{4d-orbi-co}
\end{align}

We can apply the same methods to the IR theory. 
{}In the gauge theory, we have
\begin{align}
&x = \prod_{I=1}^{2n} C_I \,, 
\quad
z = C_1 \widetilde{C}_1 = C_3 \widetilde{C}_3 = \ldots  =  C_{2n-1} \widetilde{C}_{2n-1} \,, 
\nn \\
&y = \prod_{I=1}^{2n} \widetilde{C}_I\,, 
\quad 
\tilde{z} = C_2 \widetilde{C}_2 = C_4 \widetilde{C}_4 = \ldots  =  C_{2n} \widetilde{C}_{2n}\,,
\label{T11-orbi-gauge}
\end{align}
which satisfy the algebraic equation for $T^{1,1}/\IZ_n$,  
\be
xy=(z\tilde{z})^{n} \,.
\ee
In terms of the GLSM fields in the toric diagram, the generators of the 
chiral ring are
\begin{align}
x = \prod_{k=0}^{n} (\phi_k\tilde{\phi}_k)^{n-k} \,,
\quad
y = \prod_{k=0}^n (\phi_k\tilde{\phi}_k)^k\,, 
\quad 
z  = \prod_{k=0}^n \phi_k \,,
\quad 
\tilde{z}  = \prod_{k=0}^n \tilde{\phi}_k \,.
\label{T11-orbi-glsm}
\end{align}
Comparing \eqref{T11-orbi-gauge} and  \eqref{T11-orbi-glsm}, 
we find the correspondence, 
\begin{align}
C_{2j-1} = \prod_{k=0}^{j-1} \phi_k \,,
\quad 
C_{2j} = \prod_{k=0}^{j-1} \tilde{\phi}_k \,,
\quad 
\widetilde{C}_{2j-1} = \prod_{k=j}^{n} \phi_k \,.
\quad 
\widetilde{C}_{2j} = \prod_{k=j}^{n} \tilde{\phi}_k \,.
\label{4d-orbi-co2}
\end{align}

\subsection{Orbifolding the 3d flows}

Having understood how the orbifolding works in four dimensions, 
it is straightforward to carry it over to the two flows in three dimensions 
via flavoring. 

\begin{figure}[htbp]
\begin{center}
\includegraphics[width=12cm]{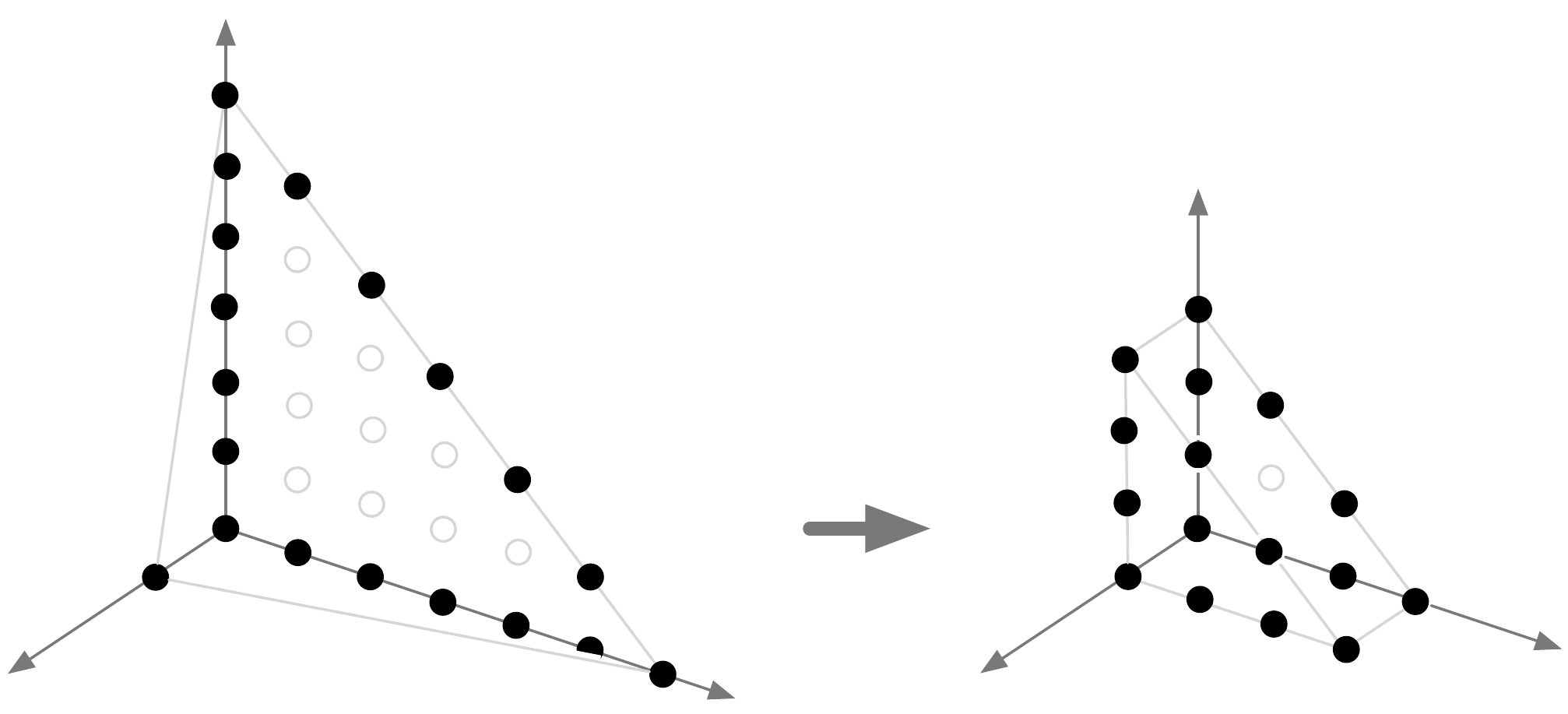}
\caption{Toric diagrams for the flow from $\IC^3/(\IZ_{2n}\times \IZ_{2n})\times \IC$ to $D_3/(\IZ_n \times \IZ_n)$
} 
\label{flow1-orbi-toric}
\end{center}
\end{figure}

\paragraph{\underline{$\IC^3/(\IZ_{2n}\times \IZ_{2n})\times \IC \goto D_3/(\IZ_n \times \IZ_n)$}} \hfill

\noindent
This flow is obtained by flavoring on all $C_i$ in the quiver diagram in Figure \ref{4dflow1-orbi-quiver}. 
In the UV theory, listing all elementary gauge invariant 
operators from the field theory and modding out by the F-term conditions, 
we find the generators of the chiral ring, 
\be
z_1 = T\,, \quad z_2 = \tilde{T}\,, \quad z_3 = \prod_{I=1}^{2n} \widetilde{C}_I\,, 
\quad 
w = C_1 \widetilde{C}_1 = \ldots  =  C_{2n} \widetilde{C}_{2n}\,.
\label{C3-orbi-gauge}
\ee
With the help of the quantum F-term relation,
\be
T \tilde{T} = \prod_{I=1}^{2n} C_I\,, 
\label{C3-orbi-qf}
\ee
we obtain the algebraic description of $\IC^3/(\IZ_{2n}\times \IZ_{2n})$,
\be
z_1 z_2 z_3 = w^{2n}, \quad v \;\; {\rm free} \,.
\label{C3-orbi-alg}
\ee
We can also see the chiral ring in the toric diagram. 
Again ignoring the trivial $\IC$ factor 
and labeling the vertices at $(0,a,b)$ by $\phi_{a,b}$, 
we find the basic gauge invariant monomials, 
\be
z_1 = \prod_{a,b} \phi_{a,b}^{2n-a-b} \,, 
\quad 
z_2 = \prod_{a,b} \phi_{a,b}^{b} \,, 
\quad 
z_3 = \prod_{a,b} \phi_{a,b}^{a} \,, 
\quad
w = \prod_{a,b} \phi_{a,b} \,. 
\label{C3-orbi-glsm}
\ee
They clearly satisfy the algebraic relation \eqref{C3-orbi-alg}. 
Comparing \eqref{C3-orbi-gauge} and \eqref{C3-orbi-glsm} 
and we find the correspondence between the gauge theory variables and the GLSM variables,
\begin{align}
C_j = \prod_{k=0}^{j-1} \hat{\phi}_k \,,
\quad 
\widetilde{C}_j = \prod_{k=j}^{2n} \hat{\phi}_k 
\quad \left( \hat{\phi}_j \equiv \prod_{b} \phi_{j,b} \right) \,.
\label{3d-orbi1-co}
\end{align}
Note the similarity between \eqref{4d-orbi-co} and \eqref{3d-orbi1-co}, 
which reflects the fact that 
the 3d theory has been obtained by the flavoring method. 

In the IR theory, the generators of the chiral ring are
\begin{align}
&z_1 = T\,, \quad z_2 = \tilde{T}\,, \quad z_3 = \prod_{I=1}^{2n} \widetilde{C}_I\,, 
\nn \\
&w = C_1 \widetilde{C}_1 = C_3 \widetilde{C}_3 = \cdots  =  C_{2n-1} \widetilde{C}_{2n-1}\,,
\nn \\
&\tilde{w} = C_2 \widetilde{C}_2 = C_4 \widetilde{C}_4 = \cdots  =  C_{2n} \widetilde{C}_{2n}\,.
\label{D3-orbi-gauge}
\end{align}
With the help of the quantum F-term relation \eqref{C3-orbi-qf},
we obtain the algebraic description of $D_3/(\IZ_{n}\times \IZ_{n})$,
\be
z_1 z_2 z_3 = (w \tilde{w})^{n}
\label{D3-orbi-alg}
\ee
In the toric diagram, we label the GLSM variables at vertices $(0,a,b)$ by $\phi_{a,b}$ and those at $(1,a,b)$ by $\tilde{\phi}_{a,b}$. 
The gauge invariant monomials are
\begin{align}
&z_1 = \prod_{a,b} (\phi_{a,b}\tilde{\phi}_{a,b})^{n-a-b} \,, 
\quad 
z_2 = \prod_{a,b}  (\phi_{a,b}\tilde{\phi}_{a,b})^{b} \,, 
\quad 
z_3 = \prod_{a,b}  (\phi_{a,b}\tilde{\phi}_{a,b})^{a} \,, 
\nn \\
&w = \prod_{a,b} \phi_{a,b} \,, 
\quad 
\tilde{w} = \prod_{a,b} \tilde{\phi}_{a,b} \,.
\label{D3-orbi-glsm}
\end{align}
They clearly satisfy the algebraic relation \eqref{D3-orbi-alg}. 
Comparing \eqref{D3-orbi-gauge} and \eqref{D3-orbi-glsm} 
and we find the correspondence between gauge theory variables and GLSM variables,
\begin{align}
&C_{2j-1} = \prod_{k=0}^{j-1} \hat{\phi}_k \,,
\quad 
C_{2j} = \prod_{k=0}^{j-1} \hat{\tilde{\phi}}_k \,,
\quad 
\widetilde{C}_{2j-1} = \prod_{k=j}^{n} \hat{\phi}_k \,.
\quad 
\widetilde{C}_{2j} = \prod_{k=j}^{n} \hat{\tilde{\phi}}_k \,, 
\label{3d-orbi1-co2}
\end{align}
where $\hat{\phi}_k$ is defined as in \eqref{C3-orbi-glsm} but 
with a different range of the $b$ index 
and similarly for $\hat{\tilde{\phi}}_k$. 
The similarity between \eqref{4d-orbi-co2} and \eqref{3d-orbi1-co2} is again a sign of the flavoring method.

\begin{figure}[htbp]
\begin{center}
\includegraphics[width=11.5cm]{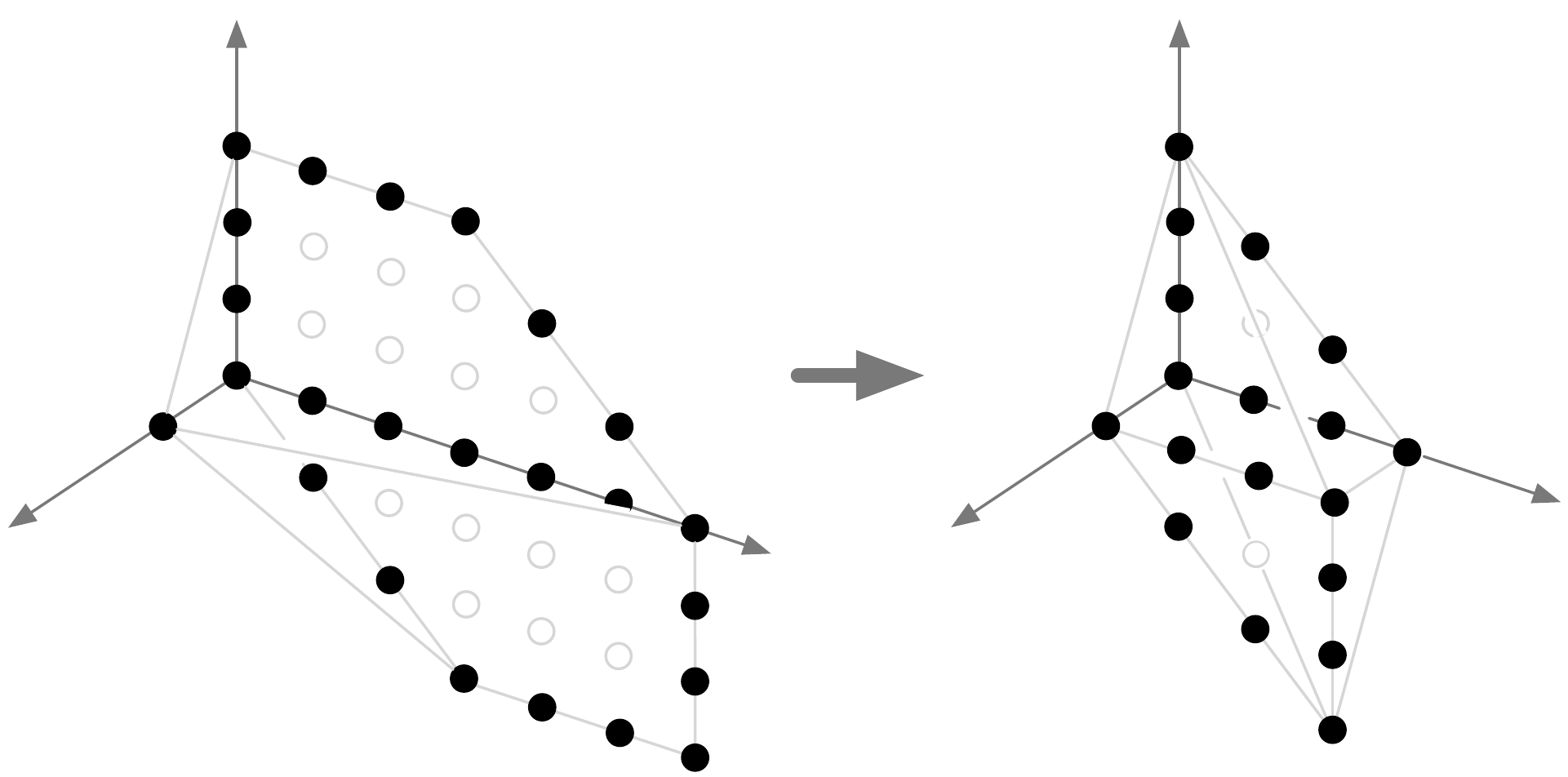}
\caption{Toric diagrams for the flow from $dP_3/(\IZ_{n}\times \IZ_{n})\times \IC$ to $Q^{1,1,1}/(\IZ_n \times \IZ_n)$
} 
\label{flow2-orbi-toric}
\end{center}
\end{figure}

\paragraph{\underline{$dP_3/(\IZ_{n}\times \IZ_{n})\times \IC \goto Q^{1,1,1}/(\IZ_n \times \IZ_n)$}} \hfill

\noindent 
This flow is obtained by flavoring on all $C_{2j-1}$ and $\widetilde{C}_{2j}$ 
in the quiver diagram in Figure \ref{4dflow1-orbi-quiver}. 
To discuss the chiral ring structure, it is convenient to introduce the (gauge-non-invariant) composite fields, 
\be
A_1 = \prod_{j=1}^n C_{2j-1}\,, \quad
A_2 = \prod_{j=1}^n \widetilde{C}_{2j}\,, \quad 
B_1 = \prod_{j=1}^n C_{2j}\,, \quad
B_2 = \prod_{j=1}^n \widetilde{C}_{2j-1}\,. \quad 
\ee
In terms of these fields, the chiral ring generators are given by 
\begin{align}
&s_1 = TB_1, \quad s_2 = \widetilde{T}B_2, \quad 
t_1 = \widetilde{T}B_1, \quad t_2 = T B_2, \quad  
u_1 = A_2 B_2, \quad u_2 = A_1 B_1,  
\nn \\
&w = C_1 \widetilde{C}_1 = \cdots  =  C_{2n} \widetilde{C}_{2n},  \quad 
v = \Phi_1 = \cdots = \Phi_{2n}.
\label{dP3-orbi-gauge}
\end{align}
Note that, for $s_1, s_2, t_1, t_2$, the gauge non-invariance of 
the composite fields cancels against that of monopole operators $T$, $\tilde{T}$. 
Using the quantum F-term relation, 
\be
T\tilde{T} = A_1 A_2 \,,
\label{dP3-orbi-qf}
\ee
we obtain the algebraic description of $dP_3/(\IZ_{n}\times \IZ_{n})\times \IC$, 
\begin{align}
&\qquad\qquad w^{2n} = s_1 s_2 = t_1 t_2 = u_1 u_2, 
\nn \\
&w^n s_1 = t_2 u_2, \quad w^n t_1 = u_2 s_2, \quad w^n u_1 = s_2 t_2,
\nn \\
&w^n s_2 = t_1 u_1, \quad w^n t_2 = u_1 s_1, \quad w^n u_2 = s_1 t_1.
\end{align}
In the toric diagram, we again ignore the trivial $\IC$ factor 
and label the vertices at $(0,a,b)$ by $\phi_{a,b}$. 
The gauge invariant monomials are
\begin{align}
&s_1 = \prod_{a,b} \phi_{a,b}^{n+b} \,, 
\quad 
t_1 = \prod_{a,b} \phi_{a,b}^{2n-a-b} \,, 
\quad 
u_1 = \prod_{a,b} \phi_{a,b}^{a} \,, 
\nn \\ 
&s_2 = \prod_{a,b} \phi_{a,b}^{n-b} \,, 
\quad 
t_2 = \prod_{a,b} \phi_{a,b}^{a+b} \,, 
\quad\quad 
u_2 = \prod_{a,b} \phi_{a,b}^{2n-a} \,, 
\quad 
w = \prod_{a,b} \phi_{a,b} \,.
\label{dP3-orbi-glsm}
\end{align}
Comparing \eqref{dP3-orbi-gauge} and \eqref{dP3-orbi-glsm}, we find the correspondence between the gauge theory variables and the GLSM variables,
\begin{align}
C_j = \prod_{k=0}^{j-1} \hat{\phi}_k \,,
\quad 
\widetilde{C}_j = \prod_{k=j}^{2n} \hat{\phi}_k 
\quad \left( \hat{\phi}_j \equiv \prod_{b} \phi_{j,b} \right) \,.
\label{3d-orbi2-co}
\end{align}
in agreement with \eqref{4d-orbi-co}. 

In the IR theory, 
the generators of the chiral ring are 
\begin{align}
&s_1 = TB_1, \quad s_2 = \widetilde{T}B_2, \quad 
t_1 = \widetilde{T}B_1, \quad t_2 = T B_2, \quad  
u_1 = A_2 B_2, \quad u_2 = A_1 B_1,  \quad
\nn \\
&w_1 = C_1 \widetilde{C}_1 = C_3 \widetilde{C}_3 = \cdots  = C_{2n-1} \widetilde{C}_{2n-1} \,, 
\quad
w_2 = C_2 \widetilde{C}_2 = C_4 \widetilde{C}_4 = \cdots   C_{2n} \widetilde{C}_{2n}\,.
\label{Q111-orbi-gauge}
\end{align}
The quantum F-term relation \eqref{dP3-orbi-qf} leads to  
the algebraic description of $Q^{1,1,1}/(\IZ_n \times \IZ_n)$,
\begin{align}
&\qquad (w_1 w_2)^n = s_1 s_2 = t_1 t_2 = u_1 u_2, 
\nn \\
&w_1^n s_1 = t_2 u_2, \quad w_1^n t_1 = u_2 s_2, \quad w_1^n u_1 = s_2 t_2,
\nn \\
&w_2^n s_2 = t_1 u_1, \quad w_2^n t_2 = u_1 s_1, \quad w_2^n u_2 = s_1 t_1.
\end{align}
In the toric diagram, we again label the GLSM variables at vertices $(0,a,b)$ by $\phi_{a,b}$ and those at $(1,a,b)$ by $\tilde{\phi}_{a,b}$. The gauge invariant monomials are
\begin{align}
&s_1 = \prod_{a,b} \phi_{a,b}^{b} \tilde{\phi}_{a,b}^{n+b} \,, 
\;
t_1 = \prod_{a,b} (\phi_{a,b}\tilde{\phi}_{a,b})^{n-a-b} \,, 
\;
u_1 = \prod_{a,b} (\phi_{a,b}\tilde{\phi}_{a,b})^{a} \,, 
\;
w_1 = \prod_{a,b} \phi_{a,b} \,,
\nn \\
&s_2 = \prod_{a,b} \phi_{a,b}^{n-b} \tilde{\phi}_{a,b}^{-b} \,, 
\;
t_2 = \prod_{a,b} (\phi_{a,b}\tilde{\phi}_{a,b})^{a+b} \,, 
\; 
u_2 = \prod_{a,b} (\phi_{a,b}\tilde{\phi}_{a,b})^{n-a} \,, 
\; 
w_2 = \prod_{a,b} \tilde{\phi}_{a,b} \,. 
\label{Q111-orbi-glsm}
\end{align}
Comparing \eqref{Q111-orbi-gauge} and \eqref{Q111-orbi-glsm} 
and we find the correspondence between the gauge theory variables and the GLSM variables,
\begin{align}
&C_{2j-1} = \prod_{k=0}^{j-1} \hat{\phi}_k \,,
\quad 
C_{2j} = \prod_{k=0}^{j-1} \hat{\tilde{\phi}}_k \,,
\quad 
\widetilde{C}_{2j-1} = \prod_{k=j}^{n} \hat{\phi}_k \,.
\quad 
\widetilde{C}_{2j} = \prod_{k=j}^{n} \hat{\tilde{\phi}}_k \,, 
\label{3d-orbi2-co2}
\end{align}
with $\hat{\phi}_k$ and $\hat{\tilde{\phi}}_k$ defined in the same way as before.


\section{Partition function and universal ratio 16/27 \label{ZF}}

Recent works have shown that the three-sphere partition function $Z$ of $\CN=2$ supersymmetric theories can be exactly computed via localization techniques. 
The exact partition function on $S^3$ provides a systematic and quantitative way 
to study the infrared dynamics 
of three-dimensional theories. In particular, it has been proposed in \cite{Jafferis:2010un} that the free energy 
$F=-\text{log}|Z|$ on $S^3$ can define a measure of the number of degrees of freedom that 
decreases monotonically along RG-flows. 
Ref.~\cite{Casini:2012ei} recently proposed a possible proof of this conjectured `F-theorem'
using a relation between the free energy $F$ on $S^3$ and a certain entanglement entropy.

In this section, we provide a more refined test of the RG flows proposed in the previous sections by computing
the exact three-sphere partition function. We will show that 
the partition functions of the UV CFT's at the UV with a certain relevant deformation exactly match those of the CFT's at the IR, which strongly supports our proposal. 

We will also observe that 
\begin{align}
  \frac{F_\text{IR}}{F_\text{UV}} = \sqrt{\frac{16}{27}}\ ,
  \label{ratiosj}
\end{align} 
for all RG flows considered in this paper, 
a result consistent with the F-theorem. We will present in section \ref{uni-proof} a general argument to explain the universal ratio $16/27$.

\subsection{Review of three-sphere partition function and large N limit}

We begin by a brief review on the three-sphere partition function of 3d $\CN=2$ supersymmetric theories and its large $N$ limit. More details can be found in \cite{Jafferis:2010un,Hama:2010av,Jafferis:2011zi} 

\paragraph{Partition function}
It has been shown in \cite{Jafferis:2010un,Hama:2010av} that the path integral for the partition function of an $\CN=2$ gauge theory coupled to charged matters on $S^3$ localizes to a matrix integral 
over the constant vev of the scalar fields in vector multiplets taking their values in the Cartan subalgebra $\mathfrak{t}$ of the given gauge group $G$. 
Explicitly, the partition function takes the following form, 
\begin{align}
  Z = \frac{1}{|W(G)|} \int_\mathfrak{t} \frac{d\l}{2\pi} \ \text{Exp} \left( \frac{i}{4\pi} \text{tr}_k \l^2 - \text{tr}_m
 \l \right) \cdot \prod_{\a\in\D} \left(2\text{sinh} \frac{\a(\l)}{2} \right) \cdot  Z_\text{1-loop}^\text{matter}(\D) \,,
\end{align}
with
\begin{align}
  Z_\text{1-loop}^\text{matter} (\D) = \prod_a \prod_{\r \in R_a} \text{Exp}\left[ l ( 1 - \D_a + i \rho(\l ) ) \right]\ ,
\end{align}
where $\D_a$ denote trial $R$-charges of matter chiral multiplets in representation $R_a$ of the gauge group $G$. Here  
$\a$ and $\r$ denote roots of the gauge group $G$ and weights of the representation $R_a$, respectively. The trace 
$\text{tr}_k$ is normalized such that, if the gauge group $G$ is given by $\prod_I G_I$ with Chern-Simons level $k_I$, 
it equals to $\sum_I k_I \text{tr}_I $. The trace $\text{tr}_m$ is normalized with the bare monopole charge $\sum_I \D^I_\text{mon} \text{tr}_I$. 
$|W(G)|$ denotes the order of the Weyl group of $G$. 
The function $l(z)$ is defined by 
\begin{align}
 l(z) = - z \text{log}(1 - e^{2\pi i z} )+ \frac i2 \Big[ \pi z^2 + \frac 1\pi
 \text{Li}_2(e^{2\pi i z}) \Big] - \frac{i\pi}{12}\ ,
\end{align}
which satisfies $ \frac{d l}{dz}(z) = - \pi z \text{cot}(\pi z)\ $ and $l(0)=0$.
Note that, since $l(+ix) + l(-ix) = 0$ for any $x \in \mathbb{R}$, the contribution from an adjoint chiral multiplet 
of $R$-charge $\D_\text{ad}=1$ becomes trivial.

\paragraph{Large $N$ limit}

Let us now discuss the general rules for evaluating the matrix integral in the large $N$ limit \cite{Herzog:2010hf,Martelli:2011qj,Cheon:2011vi}. Assuming that the eigenvalues $\l^I_i$ scale as 
\begin{align}
  \l_i^I = N^{1/2} x_i + i y^I_i + o(N^0) \ ,
\end{align}
in the large $N$ limit and replacing $x_i$ and $y_i^I$ by continuous functions $x(s)$ and $y_I(s)$ with $s\in[0,1]$,  
it turns out that the free energy functional becomes local in $x(s)$ and $y_I(s)$ for a large class of quiver gauge theories, 
including all examples in the present work. It leads to a significant simplification of the large $N$ expression of 
the free energy functional. The general rules to construct the free energy functionals in the large $N$ limit are as follows;
from now on, we restrict our attention to unitary gauge groups only.  

\begin{enumerate}
  
  \item[$\bullet$] The classical part of the action can be expressed as 
  \begin{align}
    F_\text{cl} = \frac{k_I}{2\pi} N^{3/2} \int dx \ \rho(x) x y_I(x) + \D_\text{mon}^I N^{3/2} \int dx \ \rho(x) x\ ,
  \end{align}
  where the density function $\rho(x)$ is defined as follows   
  \begin{align}
    ds = \rho\big(x(s)\big) dx(s) \ , \qquad \int dx \ \rho(x) = 1 \ .
  \end{align}
  \item[$\bullet$] Consider a pair of bifundamental matter fields, one of which in the $({\bf N}, \bar {\bf N})$  representation of $U(N)_I \times U(N)_J$ with $R$-charge $\D_{(I,J)}$ and 
  another in $(\bar {\bf N},{\bf N} )$ with $R$-charge $\D_{(J,I)}$. The contribution to the free energy functional
  from such a pair of bifundamental matter fields is given by
  \begin{align}
    F_\text{bf} = N^{3/2} \frac{2 - \D_{(I,J)}^+}{2}  \int dx \ \bigg[ \frac{\pi^2}{3} \D_{(I,J)}^+ ( 4- \D_{(I,J)}^+ ) 
   - \left( y_I - y_J + \pi \D_{(I,J)}^- \right)^2 \bigg] \ ,  
  \end{align}
  where $\D_{(I,J)}^\pm = \D_{(I,J)} \pm \D_{(J,I)}$. The above expression is valid only in the range,
  \begin{align}
    \left|y_I - y_J + \pi \D_{(I,J)}^- \right| \leq \pi \D_{(I,J)}^+\ .
  \end{align}

  \item[$\bullet$] If the gauge theory of our interest is coupled to an adjoint chiral multiplet of $R$-charge $\D_\text{ad}$, 
  the leading order contribution to the free energy functional from this multiplet becomes 
  \begin{align}
    F_\text{ad} = \frac{2\pi^2}{3} N^{3/2} \D_\text{ad} ( 1- \D_\text{ad} ) ( 2 - \D_\text{ad}) \int dx \ \rho(x)^2\ .
  \end{align}

  \item[$\bullet$] The leading order contribution to the free energy functional from a fundamental chiral multiplet of $R$-charge 
  $\D_\text{f}$ is 
  \begin{align}
    F_\text{f} = N^{3/2} \int dx \ \rho(x) |x| \left( \frac{1-\D_\text{f}}{2} - \frac{1}{4\pi} y_I(x) \right) \ .
  \end{align}
  On the other hand, the contribution from an anti-fundamental chiral multiplet of $R$-charge $\D_\text{af}$ 
  is given by 
  \begin{align}
    F_\text{af} = N^{3/2} \int dx \ \rho(x) |x| \left( \frac{1-\D_\text{af}}{2} + \frac{1}{4\pi} y_I(x) \right) \ .
  \end{align}

\end{enumerate}

\paragraph{Volume of Sasaki-Einstein sevenfold}

It has been shown in \cite{Herzog:2010hf,Jafferis:2011zi,Martelli:2011qj,Cheon:2011vi} that the free energy $F(\D)$ in the large $N$ limit as a function of trial $R$-charges $\D$ is related to the 
volume of a Sakakian seven-manifold $\text{Vol}(Y_7)[b]$ as a function of the Reeb vector $b$, provided that the gauge theory admits a gravity dual. The relation can be described as follows
\begin{align}
  F(\D) = N^{3/2} \sqrt{\frac{2\pi^6}{27 \text{Vol}(Y_7)[b]}}\ ,
  \label{ftheorem}
\end{align}
where the Reeb vector $b$ is some linear combination of the trial $R$-charges $\D$. One can determine the $R$-charges $\D_\ast$ at the fixed point via 
maximizing the free energy $F$, which translates to minimizing the Sasakian volume of $Y_7$ on the geometry side. 
At the critical value of $\D_\ast$,  the free energy $F(\D_\ast)$ is finally related to the Sasaki-Einstein volume 
$\text{Vol}(Y_7)[b_\ast]$.

\subsection{Applications to RG flows} 

The RG flows of our interest are triggered by adding a certain relevant deformation with chiral operators, i.e., 
superpotential deformation. Since the supersymmetry algebra on the three-sphere is given by  $SU(2|1) \times SU(2)$ 
containing the $U(1)_R$ symmetry, the superpotential deformation should be of $R$-charge two. This requirement introduces 
a strong constraint on the $R$-charge assignment of matter fields. However, other than this constraint, 
there will be no changes in the expression for the free energy even after such a deformation.  

We therefore expect that, given two CFT's related by the RG flows of our interest, the free energy $F_\text{IR}$ of the CFT at IR  should be the same as the free energy $F_\text{UV}$ of the CFT at UV with a restricted parameter space for the $R$-charges $\D$ due to the superpotential deformation. In general, it is not obvious that the free energy function $F_\text{UV}(\D)$ of 
an arbitrary gauge theory in three dimensions can have a critical point on such a restricted  parameter space  by  
relevant deformations. 
However, since the RG flow discussed in previous sections are proposed to have a fixed point, 
holographically dual to M-theory on Sasaki-Einstein sevenfolds, 
there must exist another critical value of $\D$  that maximizes 
the $F(\D)$ on this restricted parameter space of $R$-charges, different from the original critical value of $F_\text{UV}(\D)$ 
on the unrestricted parameter space. 

Let us now carry out the computation of the free energy for the two basic examples of RG flows discussed in section \ref{sec-gauge}.
For both examples, the partition function of the UV CFT is given by the following matrix integral
\begin{align}
  Z_\text{UV} =\frac{1}{ (N!)^2} \int \prod_{I=1}^2 \prod_{i=1}^N \frac{d\l^I_i}{2\pi} \
  \text{Exp}\left( \sum_{I,i} \D_\text{mon}^I \l^I_i \right) \cdot \text{Exp}\left( - F_\text{one-loop}\right) \ ,
  \label{Z-uv-all}
\end{align}
where $\D_\text{mon}^I$ denotes the bare $R$-charge of the monopole operators. 
The $F_\text{one-loop}$ represents
the one-loop contributions from vector multiplets and chiral multiplets, 
\begin{align}
  \text{Exp}\left( - F_\text{one-loop}\right) = \prod_{I=1}^2 \prod_{i\neq j}2\ \text{sinh}\left(
  \frac{\l_i^I - \l_j^I}{2}\right) \cdot \text{Exp}\left( - F_\text{ad} - F_\text{bf} -  F_\text{f} - F_\text{af} \right) \,.
  \label{Z-uv-1loop}
  \end{align}
The detailed form of each contribution ($F_{\rm ad}$, $F_{\rm bf}$, $F_{\rm f}$, $F_{\rm af}$) depends on the specifics of the theory. 
The RG flow is triggered by the relevant
operator 
\begin{align}
  \CW_\text{def} = \frac m2 \text{tr}\Big[ {\tilde \Phi}^2 - \Phi^2 \Big] \ .
\end{align}
which forces the $R$-charge of the adjoint fields to be $1$. This is incorporated 
in the computation of the partition function in a simple way, 
\begin{align}
  Z_\text{\rm UV} ( \D_\text{ad}=1 , \Delta_{\rm other}) = Z_{\rm IR} (\Delta_{\rm other}) \,.
\end{align}

\bigskip
\noindent \underline{$\mathbb{C}^3/(\mathbb{Z}_2 \times \mathbb{Z}_2) \times \mathbb{C} \to D_3$}

\paragraph{CFT at UV} The partition function of the orbifold CFT on the three-sphere is given by \eqref{Z-uv-all} and \eqref{Z-uv-1loop} 
with
\begin{align}
  F_{\rm ad} = & \sum_{i\neq j} l\left( 1- \D_{\Phi} + i \frac{\l_i^1 - \l_j^1}{2\pi} \right) +
  l\left( 1- \D_{\tilde \Phi} + i \frac{\l_i^2 - \l_j^2}{2\pi} \right) \ , \nonumber \\
  F_{\rm bf} = & \sum_{a=1}^2 \sum_{i,j}  l\left( 1- \D_{A_a} + i \frac{\l_i^1 - \l_j^2}{2\pi} \right)
  + l\left( 1- \D_{B_a} - i \frac{\l_i^1 - \l_j^2}{2\pi} \right) \ , \nonumber \\
  F_{\rm f} = & \sum_i l\left( 1- \D_{q} + i \frac{\l_i^2}{2\pi} \right) +
  l\left( 1- \D_{\tilde q} + i \frac{\l_i^1}{2\pi} \right)\ , \nonumber \\
  F_{\rm af} = & \sum_i l\left( 1- \D_{p} - i \frac{\l_i^1}{2\pi} \right) +
  l\left( 1- \D_{\tilde p} -i \frac{\l_i^2}{2\pi} \right)\ .
\end{align}
Due to the superpotential, one should satisfy the following relations
\begin{gather}
  \D_{\Phi} + \D_{A_1} + \D_{B_2} = 2 \ , \qquad \D_{\Phi} + \D_{A_2} + \D_{B_1} = 2\ ,
  \nonumber \\
  \D_{\tilde \Phi} + \D_{A_1} + \D_{B_2} = 2 \ , \qquad \D_{\Phi} + \D_{A_2} + \D_{B_1} = 2\ ,
  \nonumber \\
  \D_{A_1} + \D_p + \D_q = 2 \ , \qquad \D_{B_1} + \D_{\tilde p} + \D_{\tilde q}=2\ .
\end{gather}
Note that the orbifold theory has a $\mathbb{Z}_2$ flip-symmetry exchanging
\begin{align}
  A_{a} \leftrightarrow B_{a}\ ,
  \qquad \Phi \leftrightarrow \tilde \Phi\ ,
  \qquad p \leftrightarrow \tilde p \ , \qquad q \leftrightarrow \tilde q\ .
  \nonumber
\end{align}
One can therefore naturally assume that
\begin{align}
  \D_i \equiv \D_{A_i} = \D_{B_i}\ ,
  \qquad \D_\text{f} \equiv \D_p = \D_{\tilde p}\ , \qquad
  \D_\text{af} \equiv \D_q = \D_{\tilde q}\ , \qquad \D_\text{ad} \equiv \D_{\Phi} = \D_{\tilde \Phi}\ ,
  \nonumber
\end{align}
which implies that
\begin{align}
  \D_\text{ad} + \D_1 +\D_2 = 2 \ , \qquad \D_1 + \D_\text{f} +\D_\text{af} = 2\ .
\end{align}

The free energy in the large $N$ limit takes the following form
\begin{align}
  \frac{F[\r(x),y(x)]}{N^{3/2}}  
  = \int dx \ \r(x) \D_1 |x| + \int dx \ \r^2(x) (\D_1+\D_2 -2) ( y^2(x) - 4\pi^2 \D_1\D_2)\ .
\end{align}
Note that, for non-chiral theories, $\D_\text{mon} = 0$. One can show that two functions $\rho(x)$
and $y(x)$ maximizing the free energy $F$ is given by
\begin{align}
  \r(x) = \sqrt{\frac{\D_1}{8\pi^2 (2 -  \D_1-\D_2) \D_1 \D_2}}  -
  \frac{\D_1}{8\pi^2 (2 -  \D_1-\D_2) \D_1 \D_2} |x| \ , \qquad
  y(x) = 0\ ,
\end{align}
supported on $[-x_*,x_*]$ with $x_* = \sqrt{\frac{8\pi^2 (2 -  \D_1-\D_2) \D_1 \D_2}{\D_1}}$.
Plugging the above result back to $F$, one obtains
\begin{align}\label{orbtest}
  F_{\rm UV}(\D_1,\D_2) = \frac{4\pi}{3}  N^{3/2} \sqrt{2\D_1^2 \D_2 ( 2 - \D_1 -\D_2) }\ ,
\end{align}
which is maximized at
\begin{align}
  \D_1 = 1 \ , \quad  \D_2 = \frac12 \,.
\end{align}
One can read off the Sasakian volume of the orbifold space from (\ref{orbtest}),
\begin{align}
V_{\rm UV}(\Delta) =  \frac{\text{Vol}_{\rm UV}(\D)}{{\rm Vol}(S^7)} =  \frac{1}{16 \D_1^2 \D_2 ( 2 - \D_1 -\D_2)}\ ,
\end{align}
as a function of the trial $R$-charges. 
It agrees exactly with the Sasakian volume \eqref{C3-vol} of the 
$\IC^3/(\IZ_2\times \IZ_2)\times \IC$ geometry upon identification 
\begin{align}
  2\D_1 = b_2 = b_3 \ , \qquad  2\D_2 = b_1\ .
\end{align}

\paragraph{CFT at IR}

As discussed earlier, the partition function of the IR theory follows straightforwardly from that of the UV theory by $Z_\text{IR}(\Delta_1,\Delta_2) = Z_\text{UV} ( \D_\text{ad}=1,\Delta_1,\Delta_2 )$.  
The constraints due to the superpotential reads 
\begin{align}
  \D_1 +\D_2 = 1 \ , \qquad \D_1 + \D_\text{f} +\D_\text{af} = 2\ .
\end{align}
In the large $N$ limit, one can easily show that the free-energy becomes
\begin{align}
  F_{\rm IR}(\D_1) = \frac{4\pi}{3}  N^{3/2} \sqrt{2\D_1^2 ( 1- \D_1) }\ ,
\end{align}
which leads to
\begin{align}
V_{\rm IR}(\Delta) =  \frac{\text{Vol}_{\rm IR}(\D)}{{\rm Vol}(S^7)} = \frac{1}{16 \D_1^2 (1 - \D_1) }\ .
\end{align}
The above volume function exactly matches with the volume of $D_3$ \eqref{D3-vol} as a function
of the Reeb vector upon identification, 
\begin{align}
  2\D_1 = b_2=b_3 \ , \qquad b_1 =2\ .
\end{align}

\noindent \underline{$dP_3 \times \mathbb{C} \to Q^{1,1,1}$}

\paragraph{CFT at UV} The three-sphere partition function of the CFT for $dP_3 \times \mathbb{C}$ can be described 
by \eqref{Z-uv-all} and \eqref{Z-uv-1loop}
with
\begin{align}
  F_{\rm ad} = & \sum_{i\neq j} l\left( 1- \D_{\Phi} + i \frac{\l_i^1 - \l_j^1}{2\pi} \right) +
  l\left( 1- \D_{\tilde \Phi} + i \frac{\l_i^2 - \l_j^2}{2\pi} \right) \ , \nonumber \\
  F_{\rm bf} = & \sum_{a=1}^2 \sum_{i,j}  l\left( 1- \D_{A_a} + i \frac{\l_i^1 - \l_j^2}{2\pi} \right)
  + l\left( 1- \D_{B_a} - i \frac{\l_i^1 - \l_j^2}{2\pi} \right) \ , \nonumber \\
  F_{\rm f} = & \sum_i l\left( 1- \D_{q_1} + i \frac{\l_i^2}{2\pi} \right) +
  l\left( 1- \D_{q_2} + i \frac{\l_i^2}{2\pi} \right)\ , \nonumber \\
  F_{\rm af} = & \sum_i l\left( 1- \D_{p_1} - i \frac{\l_i^1}{2\pi} \right) +
  l\left( 1- \D_{p_2} -i \frac{\l_i^1}{2\pi} \right)\ .
\end{align}
Due to the superpotential, one should satisfy the following relations
\begin{gather}
  \D_{\Phi} + \D_{A_1} + \D_{B_2} = 2 \ , \qquad \D_{\Phi} + \D_{A_2} + \D_{B_1} = 2\ ,
  \nonumber \\
  \D_{\tilde \Phi} + \D_{A_1} + \D_{B_2} = 2 \ , \qquad \D_{\Phi} + \D_{A_2} + \D_{B_1} = 2\ ,
  \nonumber \\
  \D_{A_1} + \D_{p_1} + \D_{q_1} = 2 \ , \qquad \D_{A_2} + \D_{p_2} + \D_{q_2}=2\ .
\end{gather}
Note that the present model has a $\mathbb{Z}_2$ flip-symmetry exchanging
\begin{align}
  A_{1} \leftrightarrow A_{2}\ ,
  \qquad B_{1} \leftrightarrow B_{2}\, \qquad \Phi \leftrightarrow - \Phi\ ,
  \qquad \tilde \Phi \leftrightarrow - \tilde \Phi\ ,
  \qquad p_1 \leftrightarrow p_2 \ , \qquad q_1 \leftrightarrow q_2\ . \nonumber
\end{align}
One can therefore naturally assume that
\begin{align}
  \D_A \equiv \D_{A_i}\ , \qquad \D_B \equiv \D_{B_i}\ ,
  \qquad \D_p \equiv \D_{p_i} \ , \qquad \D_q \equiv  \D_{q_i}\ ,
\end{align}
and then
\begin{align}
  \D_{ad} = \D_\Phi = \D_{\tilde \Phi}\ , \qquad  \D_{ad} + \D_A +\D_B = 2 \ , \qquad \D_A + \D_p +\D_q = 2\ .
\end{align}

In the large $N$ limit, the free energy  is given by
\begin{align}
  \frac{F[\r(x),y(x)]}{N^{3/2}} & =  
  \D_{\rm mon} \int dx\ \r(x) x + \int dx \ \r(x) |x| \left(\D_A + \frac{1}{2\pi} y(x) \right)^2 
  \nonumber \\ & \ \ \ 
  + (\D_A + \D_B -2 ) \int dx \ \r^2(x) \left( y(x) + 2\pi \D_A \right)\left( y(x) - 2\pi \D_B \right) \ ,
\end{align}
where $y(x)=y_1(x) -y_2(x)$. The above expression is valid only in the region 
\begin{align}\label{validitydp3}
  \left| \frac{y(x)}{\pi} + \D_A - \D_B \right| \leq \D_A + \D_B\ .
\end{align}
One can show that two functions $\rho(x)$ and $y(x)$ maximizing the free energy $F$ 
are given by
\begin{align}
  \r(x) & = \frac{(\D_A +\D_B) |x| + 2 (x \D_{\rm mon} + \m)}{4\pi^2 ( \D_A +\D_B) (\D_A+\D_B -2)}
  \nonumber \\ 
  y(x) & = -2\pi \frac{(\D_A-\D_B)(x\D_{\rm mon} + \m) + \D_A(\D_A+\D_B) |x|}{2(x\D_{\rm mon} +\m) + (\D_A+\D_B) |x|}\ .
\end{align}
Due to the condition (\ref{validitydp3}), the density functions are supported on $[x_l,x_r]$ where
\begin{align}
  x_l = \frac{\m}{\D_A+\D_B- \D_{\rm mon}} \ , \qquad x_r = -\frac{\m}{\D_A+\D_B + \D_{\rm mon}}\ ,
\end{align}
and
\begin{align}
   \m = - 2\pi \left((\D_A +\D_B)^2- \D^2 \right)
  \sqrt{\frac{(\D_A+\D_B) ( 2 - \D_A -\D_B)}{3 (\D_A +\D_B)^2 - \D_{\rm mon}^2}}\ . 
\end{align}
Plugging the above result back to the free energy functional $F$, one obtains
\begin{align}\label{dP3test}
  \frac{F}{N^{3/2}} = \frac{4\pi}{3} (\D_A+\D_B) \left( (\D_A+\D_B)^2 - \D_\text{mon}^2\right)
  \sqrt{\frac{2 - \D_A -\D_B}{\left( 3(\D_A+\D_B)^2 - \D_{\rm mon}^2 \right) ( \D_A+\D_B)}}\ ,
\end{align}
which is maximized at
\begin{align}
  \D_{\rm mon} = 0 \ , \ \D_A+\D_B = \frac32\ .
\end{align}
As discussed in \cite{Jafferis:2011zi}, one cannot determine the $R$-charge $\D_A$ or $\D_B$ for $U(N)\times U(N)$ theory due to the flat direction 
of the partition function. For $SU(N)\times SU(N)\times U(1)$ theory where the dibaryon operators are allowed, an extra condition can be used to determine $\D_A$: 
\begin{align}
  \int dx \, \r(x)y(x) = 0 \ \to \ \D_A = 1 \ , \ \ \D_B= \frac12\ .
\end{align}
At the critical value $\D_A=1$ and $2\D_B=1$, one finally confirms 
that the free energy $F$ reproduces the 
volume of the base of the cone $dP_3 \times \mathbb{C}$,  
\begin{align}
F_{\rm UV} \;\; \goto \;\; V_{dP_3\times \mathbb{C}} = \frac{{\rm Vol}(dP_3\times \IC)}{{\rm Vol}(S^7)} = \frac{2}{9} \,.
\end{align}

\paragraph{CFT at IR} The partition function of the $Q^{1,1,1}$ theory perfectly agrees with that of the $dP_3\times\mathbb{C}$ 
theory with $\D_\text{ad} = 1$, i.e., 
\begin{align}
  Z_{dP_3\times\mathbb{C} } ( \D_\text{ad}=1 ) = Z_{Q^{1,1,1}} 
\end{align}
with
\begin{align}
  \D_A +\D_B = 1 \ , \qquad \D_A + \D_p +\D_q = 2\ .
\end{align}

The large $N$ limit of the free energy now becomes
\begin{align}
  \frac{F[\r(x),y(x)]}{N^{3/2}} & =  
  \D_\text{mon} \int dx\ \r(x) x + \int dx \ \r(x) |x| \left(\D_A + \frac{1}{2\pi} y(x) \right)^2
  \nonumber \\ & \ \ \
  - \int dx \ \r^2(x) \left( y(x) + 2\pi \D_A \right)\left( y(x) + 2\pi (\D_A -1 )  \right) \ ,
\end{align}
where $y(x)=y_1(x) -y_2(x)$. The above expression is valid only in the region 
\begin{align}\label{validityQ111}
  \left| \frac{y(x)}{\pi} + 2\D_A - 1 \right| \leq 1\ .
\end{align}
One can show that two functions $\rho(x)$ and $y(x)$ maximizing the free energy $F$ 
are given by 
\begin{align}
  \r(x) & = - \frac{|x| + 2 (x \D_\text{mon} + \m)}{4\pi^2}
  \nonumber \\
  y(x) & = -2\pi \frac{(2\D_A-1)(x\D_\text{mon} + \m) + \D_A |x|}{2(x\D_{\rm mon} +\m) + |x|}\ .
\end{align}
Due to the condition (\ref{validityQ111}), the density functions are supported on $[x_l,x_r]$ where
\begin{align}
  x_l = \frac{\m}{1- \D_\text{mon}} \ , \qquad x_r = -\frac{\m}{1+ \D_\text{mon}}\ , 
  \qquad
   \m = - \frac{2\pi \left(1 - \D^2 \right)}
  {\sqrt{3  - \D_{\rm mon}^2}}\ .
\end{align}
Plugging the above result back to $F$, one obtains
\begin{align}\label{Q111test}
  \frac{F}{N^{3/2}} = \frac{4\pi}{3}  \left(1  - \D_\text{mon}^2\right)
  \sqrt{\frac{1}{3 - \D_\text{mon}^2 }}\ ,
\end{align}
which is maximized at
\begin{align}
  \D_\text{mon} = 0 \ .
\end{align}
Again, one cannot determine the $R$-charge $\D_A$ or $\D_B$ for $U(N)\times U(N)$ theory due to the flat direction
of the partition function \cite{Jafferis:2011zi}. For $SU(N)\times SU(N)\times U(1)$ theory where the dibaryon
operators are allowed, the condition below determines $\D_A$:
\begin{align}
  \int dx \ \r(x)y(x) = 0 \ \to \ \D_A = \frac23 \ , \ \ \D_B= \frac13\ .
\end{align}
At this critical value $\D_A=2/3$ and $\D_B=1/3$, 
one finally confirms 
that the free energy $F$ reproduces the 
volume of $Q^{1,1,1}$,  
\begin{align}
F_{\rm IR} \;\; \goto \;\; V_{Q^{1,1,1}} = \frac{{\rm Vol}(Q^{1,1,1})}{{\rm Vol}(S^7)} = \frac{3}{8} \,.
\end{align}

\subsection{Field theoretical proof of universal ratio \label{uni-proof}}

We consider a slightly broader generalization of the `necklace' quiver gauge theory considered in section \ref{sec-orbi}. The gauge theory of our interest has gauge group $G=\prod_{I=1}^L U(N)_I$ with Chern-Simons levels $k_I$ subject to $\sum_I k_I = 0$. Turning on Chern-Simons levels leads to a straightforward generalization 
of the RG flows discussed previous sections. While we do not intend to repeat 
the the analyses of previous sections for this larger class of theories, we will include 
the possibility of adding Chern-Simons levels in the computation of the partition function in this subsection. 

The gauge fields couple to $U(N)_I \times U(N)_{I+1}$ bifundamental and anti-bifundamental chiral multiplets $C_I, \widetilde{C}_I$ ($I=1,2,\ldots,L$)
and one adjoint chiral multiplet $\Phi_I$ for each $U(N)_I$. 
The superpotential of the UV theory prior to flavoring is the same as before, 
\begin{align}
  \CW^{(0)}_{{\rm UV}} = \sum_{I=1}^L \text{tr} \left( \Phi_I C_I \widetilde{C}_I - \Phi_I \widetilde{C}_{I-1} C_{I-1} \right) \,.
\end{align}
Next, we consider arbitrary flavoring by coupling the quiver theory with two types of paired chiral multiplets, denoted by 
$(p_I^a, q_{I+1}^a)$ ($a=1,2,\ldots,n_{I}$) and $(\tilde p_i^{\dot{a}} , \tilde q_{I+1}^{\dot{a}})$ ($\dot{a}=1,2,\ldots,\tilde{n}_I$),  
which introduces the superpotential term
\begin{align}
  \CW_\text{flavor}= \sum_{I=1}^L \left[ \sum_{a=1}^{n_I} \text{tr} \left( p_I^a C_I q_{I+1}^a\right) 
  + \sum_{\dot{a}=1}^{\tilde{n}_I} \text{tr} \left( \tilde{q}_{I+1}^{\dot{a}} \widetilde{C}_{I+1} \tilde{p}_I^{\dot{a}} \right) \right]\ .
  \label{W-flavor-general}
\end{align}
After the RG flow, the IR theory takes the bare superpotential
\be
\CW_{\rm IR}^{(0)} = \frac{1}{m} \sum_{I=1}^{L} (-1)^{I} \tr\left( C_I \widetilde{C}_I \widetilde{C}_{I-1} C_{I-1} \right) \,, 
\ee
and shares the same flavoring superpotential \eqref{W-flavor-general}.

The free energy of the UV theory in the large $N$ limit now takes the following form 
\begin{align}
  \frac{F}{N^{3/2}} & =  \sum_I \frac{k_I}{2\pi} \int dx\ \rho(x) x y_I(x) + \D_{\rm mon} \int dx \rho(x) x 
\nn \\  
& + \frac{2\pi^2}{3} \sum_I \D_{\Phi_I} ( 1 - \D_{\Phi_I} ) (2 -\D_{\Phi_I} ) \int dx\ \rho^2(x)  
  \nonumber \\ &
  + \sum_I \frac{2 - \D_{C_I}- \D_{\widetilde{C}_I}}{2} \int dx\ \rho^2(x) \left[ 
  \frac{\pi^2}{3} (\D_{C_I} + \D_{\widetilde{C}_I} ) ( 4- \D_{C_I}- \D_{\widetilde{C}_I}) \right.
  \nonumber \\ & \hspace*{6.3cm} \left.
  - \big(y_I(x) - y_{I+1}(x) + \pi (\D_{C_I}- \D_{\widetilde{C}_I}) \big)^2 \right]
  \nonumber \\ &
  + \sum_{I} \sum_a \int dx \ \rho(x) |x| \left( \frac{2-\D_{p^a_I}-\D_{q^a_{I+1}}}{2} 
  + \frac{1}{4\pi} \big( y_I(x)-y_{I+1}(x) \big)  \right) 
  \nonumber \\ & 
  + \sum_{I} \sum_{\dot{a}} \int dx \ \rho(x) |x| \left( \frac{2-\D_{\tilde p^{\dot{a}}_I}-\D_{\tilde q^{\dot{a}}_{I+1}}}{2}
  - \frac{1}{4\pi} \big( y_I(x)-y_{I+1}(x) \big)  \right) \ , 
  \label{F-general}
\end{align}
which is valid when the following conditions are satisfied for all $I$
\begin{align}
  \Big| y_I - y_{I+1} + \pi\big( \D_{C_I} - \D_{\widetilde{C}_I} \big) \Big|
  \leq \pi \big( \D_{C_I} + \D_{\widetilde{C}_I} \big)\ ,
  \label{validitygeneral}
\end{align}
and the density function $\rho(x)$ should satisfy the relation below
\begin{align}
  \int_{\mathfrak R} dx \  \rho(x) = 1\ .
  \label{densitygen}
\end{align}

In order for the UV theory to be a CFT, all monomials in the superpotential 
must have $R$-charge 2. It follows that
\begin{align}
  &\D_{\Phi} = \D_{\Phi_I} \ \text{for all}\  I\ , \qquad\qquad\qquad\qquad
  \D_{\Phi} + \D_{C_I} + \D_{\widetilde{C}_I} = 2 \ \text{for all}\ I \ , 
  \label{relation5}
  \\
  &\D_{C_I}+ \D_{p^a_I} + \D_{q^a_{I+1}} = 2 \ \text{for all }\ a\ , 
  \qquad
  \D_{\widetilde{C}_I}+ \D_{\tilde p^{\dot{a}}_I} + \D_{\tilde q^{\dot{a}}_{I+1}} = 2 \ \text{for all}\ \dot{a}\ .
  \label{relation5b}
\end{align}
Using \eqref{relation5b}, we can eliminate all dependences on $(\D_{p^a_I}, \D_{q^a_{I+1}}, \D_{\tilde p^{\dot{a}}_I}, \D_{\tilde q^{\dot{a}}_{I+1}})$ from 
\eqref{F-general}, so that $F$ becomes a function of $(\Delta_\Phi, \Delta_{C_I}, \Delta_{\widetilde{C}_I})$ only. 
For notational simplicity, we will write $(\Delta_I, \widetilde{\Delta}_I)$ 
to denote $(\Delta_{C_I}, \Delta_{\widetilde{C}_I})$ 
in what follows. 

One can rearrange the above free energy into the following form
\begin{align}
  \frac{F}{N^{3/2}}  =  F_\text{linear} + F_\text{quadratic} 
\end{align}  
with
\begin{align}
  F_\text{linear} & = 
  \sum_I \frac{k_I}{2\pi} \int dx\ \rho(x) \,x\, y_I(x) + \D_{\rm mon} \int dx \rho(x) x
  \nonumber \\  &
  + \sum_{I} n_{I} \int dx \ \rho(x) |x| \left[ \frac{\D_I}{2}
  + \frac{1}{4\pi} \big( y_I(x)-y_{I+1}(x) \big)  \right]
  \nonumber \\ &
  + \sum_{I} \tilde n_{I} \int dx \ \rho(x) |x| \left[ \frac{\widetilde{\Delta}_I}{2}
  - \frac{1}{4\pi} \big( y_I(x)-y_{I+1}(x) \big)  \right] \ . 
\end{align}
and 
\begin{align}
  F_\text{quadratic} =  \sum_I \frac{2 - \D_I- \widetilde{\D}_I}{2} \int dx\ \rho^2(x)
  & \left[ 2\pi \D_{i} + \big(y_I(x) - y_{I+1}(x) \big) \right] \nonumber 
  \\ \times &   \left[ 2\pi \widetilde{\D}_I - \big(y_I(x) - y_{I+1}(x) \big) \right] \ .
\end{align}

It is easy to show that the above CFT can flow down to another CFT at the infrared  
once we turn on complex mass terms for adjoint chiral multiplets. 
In order to relate the partition function of the UV CFT to that of the IR CFT, 
it is useful to rescale the parameters entering the partition function: 
\begin{itemize}

  \item Rescale the conformal dimensions of bifundamental matters as follows 
  \begin{align}
    \D'_I \equiv \frac{\D_I}{\D_{I}+\widetilde{\D}_I} =
    \frac{\D_I}{2 - \D_\Phi } \ , 
    \quad 
    \widetilde{\D}'_I \equiv \frac{\widetilde{\D}_I}{\D_{I}+\widetilde{\D}_I} =
    \frac{\widetilde{\D}_I}{2 - \D_\Phi } \ , 
    \label{rescaling}
  \end{align}
  which guarantees, for all $I$,
  \begin{align}
    \D'_I + \widetilde{\D}'_I = 1 \ .
    \label{relation6}
  \end{align}
  The rescaled variables $(\D'_I, \widetilde{\D}'_I)$ will later be identified with 
  the $R$-charges of the CFT at IR, where the marginality of the superpotential will require \eqref{relation6}. 
  
  \item Rescale functions $y_I(x)$ as follows
  \begin{align}
    \hat{y}_I (\hat x) \equiv \frac{y_I(x)}{\D_{I}+\widetilde{\D}_I} = 
    \frac{y_I(x)}{2 - \D_\Phi }\ , \qquad \hat x = \alpha x\ .
  \end{align}
  The value of the constant $\alpha$ will be determined later. After this rescaling, (\ref{validitygeneral}) implies 
  another relation one need to satisfy for the CFT at IR
  \begin{align}
    \left| \hat y_I(\hat x) + \pi \left( \D'_{I}-\widetilde{\D}'_I\right) \right|
    \leq \pi\ .
  \end{align}
  It also guarantees that the rescaled density function $\hat \r(\hat x)$, defined below, 
  is supported on the rescaled region $\widehat {\mathfrak R}$. 
  
  \item Rescale the density function $\rho(x)$ as follows 
  \begin{align}
    \beta \hat \rho (\hat x) = \rho ( x) \ ,    
  \end{align}
  where $\beta$ is a constant. One can fix this constant $\beta$ in terms of $\alpha$ by requiring that 
  \begin{align}
    \int_{\hat {\mathfrak R}}d\hat x \hat \r(\hat x) =1\ , 
  \end{align}
  which is again needed for the CFT at IR. From the relation (\ref{densitygen}), one can show that
  \begin{align}
    1 = \int_\mathfrak{R} dx\ \rho(x) = \alpha \beta \int_{\widehat {\mathfrak R}}d\hat x \hat \r(\hat x)
    \ \imp \ \alpha \beta =1\ .
  \end{align}

  \item Under the above rescaling, the free energy of the CFT at UV scales as 
  \begin{align}
    F_\text{linear} & =  \alpha (2 - \D_{\Phi} ) \widehat F_\text{linear}(\D'_{I},\widetilde{\D}'_I)\ , 
    \nonumber \\ 
    F_\text{quadratic}&  =  \alpha^{-1} (2 - \D_{\Phi} )^2 \D_{\Phi} 
    \widehat F_\text{quadratic}(\D'_{I},\widetilde{\D}'_I)\ .
  \end{align} 
  Choosing the constant $\alpha$ by 
  \begin{align}
    \alpha = \sqrt{\D_{\Phi} \big(2 - \D_{\Phi}\big)}, 
  \end{align}
  one finally obtains
  \begin{align}
    F_\text{UV} (\D_\Phi, \D_I, \widetilde{\D}_I) = & \sqrt{\D_{\Phi} \big(2 - \D_{\Phi}\big)^3} \times 
    \Big(  \widehat F_\text{linear}(\D'_{I},\widetilde{\D}'_I) 
    + \widehat F_\text{quadratic}(\D'_{I},\widetilde{\D}'_I) 
    \Big) \nonumber \\ = &  
    \sqrt{\D_{\Phi} \big(2 - \D_{\Phi}\big)^3} \times  
    F_\text{IR}(\D'_{I},\widetilde{\D}'_I)\ ,
    \label{relationUVIR}
  \end{align}
  where $F_\text{IR}$ can be identified as the free energy of the CFT at IR.

\end{itemize} 
From (\ref{relationUVIR}), one can show that the free-energy of the CFT at UV should be 
extremized at 
\begin{align}
  \D_\Phi = \frac12\ , 
\end{align}
which leads to the very universal ratio between the free energies at IR and UV:
\begin{align}
  F_\text{UV} (\D_\Phi, \D_I, \widetilde{\D}_I) 
  = \sqrt{\frac{27}{16}} F_\text{IR}(\D'_{I},\widetilde{\D}'_I)\ .
\end{align}

A few final remarks are in order. First, recall that the relevant term 
in the superpotential \eqref{W-def-quiver}
which triggers the KW flow have a finely tuned 
relative ratio of mass-deformation parameters.  
However, the proof given above on the universal ratio $16/27$  
are rather insensitive to these ratios. It strongly suggests that there should exist a manifold of RG fixed points continuously connected to the KW flow. 
Second, we can slightly modify the above proof to incorporate other flows such as the PW flows. 
For instance, consider the RG flows induced by a cubic superpotential for adjoint chiral multiplets. For such a 
deformation, $\D_\Phi = 2/3$. We therefore need to choose a different rescaling for conformal dimensions, 
\be
 \D_I' = \frac{4 \D_I}{3 (2 - \D_\Phi )}\ , \qquad 
\tilde \D_I' = \frac{4 \D_I}{3 (2 - \D_\Phi )}\ , \qquad 
\D_I' + \tilde \D_I' = \frac43\ .
\ee
The rest of the proof can proceed in parallel with the one shown above, leading to another 
universal ratio 
\be
F_\text{UV} = \frac{32\sqrt2}{27\sqrt3}F_\text{IR} \,.
\ee
The cubic deformation and its universal ratio $2^{11}/3^{7}$ was first observed again in \cite{Jafferis:2011zi}, and its supergravity solutions were discussed in a recent work \cite{Gabella:2012rc}.

\section{Discussion}


In this paper, we argued for the existence of two KW flows in M-theory 
and their generalization to orbifolds. 
All examples we considered in this paper, 
and further generalization with non-vanishing Chern-Simons levels, share some common features. 
The gauge theories are constructed by applying 
the flavoring method to dimensionally reduced KW flow. 
Geometrically, the flavoring lifts the toric diagram of a CY$_3$ to 
that of a CY$_4$.  

A natural open question is whether there are more flows of KW type 
that are not in the same family as those considered in this paper. 
One possible approach would be to find a new KW-like flow in four dimensions
(other than the orbifold generalization) and apply the flavoring method. 
In fact, that seems to be essentially the only viable approach one can 
take because all known methods to construct $2\le \CN \le 6$ Chern-Simons-matter 
theories with large $N$ AdS dual involves KK reduction in one way or another. 

For M2-branes probing a CY$_4$ cone, a genuinely three dimensional 
approach to construct the CFT was proposed in \cite{crystal1,crystal2,crystal3}. 
Being a three dimensional analog of the brane tiling model for four dimensional CFT's \cite{Franco:2005sm,Hanany:2005ss}, it was later called a brane crystal model.
The model was not sufficiently developed to fully specify 
the associated CFT for multiple M2 branes. But, in the abelian case, it did give some non-trivial information on the matter contents, superpotential terms 
and gauge groups, and successfully reproduced the chiral ring and the spectrum of baryons. 

The brane crystal model can be used to study RG flows in M-theory. 
Ref.~\cite{crystal3} gave a diagrammatic understanding of the original KW flow in terms of the brane tiling model and, by extending the idea to the brane crystal model, it predicted the existence of the new KW flows in M theory 
discussed in this paper. It would be interesting to 
improve the brane crystal model by using some of the recent developments 
in M2-brane CFT's and use it to search for more new flows.

In the brane crystal model, the connection 
between the three dimensional gauge theory and its four dimensional counterpart 
can be understood as the projection of the crystal diagram onto a tiling diagram. 
There are two qualitatively different types of projections. There are relatively few cases where the projection 
can be done without double lines or crossings. There are many more cases 
where double lines or crossings are unavoidable. 
In the former case, it was shown in \cite{Imamura:2008qs} that 
the Chern-Simons levels can encode the information on the `vertical' moves 
of the crystal diagram. In the more subtle latter case, the flavoring method of \cite{Aganagic:2009zk,Benini:2009qs} was needed to account for the vertical moves. 
Some crystal diagrams can be projected down to two or more inequivalent tiling diagrams, leading to different field theory Lagrangians. It would be interesting 
to test the quantum equivalence among different projections by, for instance, 
computing the three-sphere partition function.

Finally, it would be interesting to find an explicit supergravity solution 
describing the new KW flow in M-theory, perhaps with the help of the recently 
developed methods for analyzing AdS$_4$ solutions in M-theory  \cite{Gabella:2011sg,Gabella:2012rc} and related works 
for the original KW flow in four dimensions \cite{Halmagyi:2004jy}.

\begin{acknowledgments}
%
We would like to thank Francesco Benini and Stefano Cremonesi for explanations of their work.   
We also thank the organizer of the workshops  
{\it String Theory}, July 2011 at the Centro de Ciencias de Benasque, Spain 
and 
{\it Mathematics and Applications of Branes in String and M-theory}, 
2012 at the Isaac Newton Institute for Mathematical Sciences, 
Cambridge, UK for hospitality.
The work of Sangmin Lee is supported by 
National Research Foundation of Korea (NRF) Grants  
2009-0072755, 2009-0084601 and 2012R1A1B3001085. 
The work of Sungjay Lee is supported by the Ernest Rutherford fellowship of the Science \& Technology Facilities Council ST/J003549/1.

\end{acknowledgments}

\end{document}